\documentclass[final,authoryear,5p]{elsarticle}

\usepackage{epsfig}

\usepackage{amssymb}

\usepackage[
a4paper=true,%
breaklinks=true,%
colorlinks=true,%
pdfauthor={First Author et al.},%
pdftitle={Template for manuscripts in Advances in Space Research}%
]{hyperref}

\usepackage{graphicx}
\usepackage{color} 
\usepackage{natbib}

\journal{Advances in Space Research}
\usepackage{textcomp}

\begin{document}

\begin{frontmatter}

\title{Scaling Laws of Quiet-Sun Coronal Loops.}





\author[iafe,untref]{C. Mac Cormack}
\ead{cmaccormack@iafe.uba.ar}
\author[iafe]{M. L\'opez Fuentes}
\author[iafe,uba]{C.H. Mandrini}
\author[iafe]{D. G. Lloveras}
\author[iafe,untref]{M. Poisson}
\author[iafe,untref]{A.M. V\'asquez}

\address[iafe]{Instituto de Astronom\'ia y F\'isica del Espacio (IAFE, UBA-CONICET), Ciudad Aut\'onoma de Buenos Aires, Argentina.}
\address[untref]{Departamento de Ciencia y Tecnolog\'ia, Universidad Nacional de Tres de Febrero, Caseros, Provincia de Buenos Aires, Argentina.}
\address[uba]{Departamento de F\'isica, Universidad de Buenos Aires (UBA), Ciudad Aut\'onoma de Buenos Aires, Argentina.}

\begin{abstract}

We study a series of relations between physical parameters in coronal loops of the quiet Sun reconstructed by combining tomographic techniques and modeling of the coronal magnetic field. We use differential emission measure tomography (DEMT) to determine the three-dimensional distribution of the electron density and temperature in the corona, and we model the magnetic field with a potential-field source-surface (PFSS) extrapolation of a synoptic magnetogram. By tracing the DEMT products along the extrapolated magnetic field lines, we obtain loop-averaged electron density and temperature. Also, loop-integrated energy-related quantities are computed for each closed magnetic field line. We apply the procedure to Carrington rotation 2082, during the activity minimum between Solar Cycles 23 and 24, using data from the {Extreme Ultraviolet Imager} on board the { Solar Terrestrial Relations Observatory} (STEREO) spacecraft. We find a scaling law between the loop-average density $N$ and loop length $L$, $N_m \sim L^{-0.35}$, but we do not find a significant relation between loop-average temperature and loop length. We confirm though the previously found result that loop-average temperatures at the equatorial latitudes are lower than at higher latitudes. We associate this behavior with the presence at the equatorial latitudes of loops with decreasing temperatures along their length (``down'' loops), which are in general colder than loops with increasing temperatures (``up'' loops). We also discuss the role of ``down'' loops in the obtained scaling laws of heating flux {\it versus} loop length for different heliographic latitudes. We find that the obtained scalings for quiet-Sun loops do not generally agree with those found in the case of AR loops from previous observational and theoretical studies. We suggest that to better understand the relations found, it is necessary to forward model the reconstructed loops using hydrodynamic codes working under the physical conditions of the quiet-Sun corona.
\end{abstract}


\end{frontmatter}


\section{Introduction}
\label{intro}

The study of the energy balance of the magnetically closed corona provides clues on the mechanisms that maintain this region of the solar atmosphere {two} orders of magnitude hotter than the photosphere. The magnetized and rarefied coronal plasma imply that transport phenomena are strongly inhibited in directions perpendicular to the magnetic field lines. This is particularly evident in active regions (ARs), in which the plasma is structured in the form of loops and arcades as observed in extreme-ultraviolet (EUV) and soft X-ray (SXR) images obtained with space telescopes. Loops can be thought then, as individual one-dimensional (1D) atmospheres with more or less independent evolutions. 

Seminal studies \citep{rosner_1978, vesecky_1979} {found that X-ray observations are consistent with loops in equilibrium, assuming energy balance between heating and thermal and radiative losses. Several scaling laws deduced from these studies were used to determine whether loops are in equilibrium or not. For instance, \citet{vesecky_1979} found that in quasi-static equilibrium the three terms of the balance equation should be approximately equal. It can be easily demonstrated \citep[see, e. g.,][]{lopezfuentes_2007} that this condition implies a scaling law between density and temperature, $N \approx T^2$. However, these results are based on X-ray loops with temperatures well above 2 MK \citep[usually called "hot loops", see, e. g.,][]{reale_2014}. Later analysis based on EUV data, showed that warm loops (with temperatures around 1.2 MK) are too dense to be in static or quasi-static equilibrium \citep{aschwanden_2001,winebarger_2003}. All the mentioned results correspond to observations of AR loops.} 

Regarding the energy input needed to maintain observed coronal conditions, the review by \citet{withbroe_1977} provided estimations based on phenomenological models of the order of 10$^7$ erg cm$^{-2}$ s$^{-1}$ for ARs and 3$\times10^5$ erg cm$^{-2}$ s$^{-1}$ for the quiet Sun {(also see, \citet{hahn_2014,maccormack_2017})}. These estimations are still used as canonical values. 

One way to test different coronal heating models has been to study the presence of scaling laws between observed physical parameters of ARs coronal loops, such as temperature, density, length and magnetic strength, and compare them with the predictions of the models \citep[see, e.g.,][]{fisher_1998, mandrini_2000,jain_2006,pevtsov_2003}. 

Since magnetic loops are the basic observable blocks of the coronal structure and they are most conspicuously observed in ARs, most efforts to understand the energy input and coronal dynamics have been focused on studying AR coronal loops \citep[see, e.g., the reviews by ][]{reale_2014, klimchuk_2015}. Due to a lower intensity and apparent uniformity in EUV and SXR observations, the situation is quite different in the quiet-Sun corona, where a direct identification of loops is normally not possible. For this reason, quiet-Sun studies have been much scarce. In the case of quiet-Sun loops, it is necessary to reconstruct the location of magnetic field lines and the plasma parameters along them by other means \citep[see, e.g.,][]{hahn_2014}.

In this work, we use differential emission measure tomography (DEMT) along a full solar rotation, to reconstruct the 3D distribution of the temperature and electron density in the coronal volume between 1.02 and 1.225 $R_{\odot}$ \citep{vasquez_2016}. {Since we use one EUV image every six hours to obtain a global description of the corona, the tomographic technique does not resolve the shorter timescales of the coronal dynamics. Furthermore, given the height limits imposed on the tomography, low-lying structures such as coronal bright points are left out of the analysis. Thus, in this paper, we only study the large scale Quiet Sun corona.} The DEMT results are combined with a potential magnetic field model (PFSS) that extrapolates the global solar magnetic field from observed synoptic magnetograms. In this way, we obtain the loop-average temperature and density of the plasma in each closed field line integrated from the magnetic model. Also, we find a loop-integrated equation derived from the energy balance between a heating term, radiative cooling and conductive flux (Mac Cormack et al., 2017). This allows derivation of the energy input flux required at the coronal base to maintain thermodynamically stable coronal structures. With the objective of finding statistical relations and identifying possible scaling laws between observed and inferred coronal parameters, we apply this procedure to the particular case of Carrington rotation 2082, that occurred at the minimum between Solar Cycles 23 and 24 with no relevant ARs on the Sun during that rotation.

In Section~\ref{MethodandData} we present a description of the tomographic technique, the potential model and the energy balance model used to describe energy fluxes along reconstructed loops, as well as the data used and the methodology followed. In Section~\ref{Results} and its subsections we present and analyze our results in terms of the studied loop parameters and we discuss and conclude in Section~\ref{Conclusion}.\\

\section{Method and Data}
\label{MethodandData}

\subsection{Tomographic Technique and PFSS Model}
\label{Tomo}

The {differential emission measure tomography} (DEMT) is a technique developed by \citet{frazin_2009} and used in several works focused on the study of coronal plasma properties during minima of solar activity cycles (\citealp{nuevo_2015,lloveras_2017}; and the review by \citealp{vasquez_2016}). The technique divides the corona in a spherical grid that covers all latitudes and longitudes between {1.025 $R_{\odot}$ and 1.225 $R_{\odot}$}. Using a series of EUV images from different filter bands of the instrument used, covering a full solar rotation, it obtains a 3D distribution of the emissivity in the coronal volume of interest. The tomographic emissivities in each computational cell are then used to determine its local differential emission measure (LDEM), which describes the temperature distribution of the electron plasma within the specific voxel. The method relies on a parametric-modeling technique, finding the LDEM that best predicts the tomographic data. 

Previous works \citep{nuevo_2015} showed that LDEMs can be modeled using Gaussian functions determined by three parameters: area, mean temperature, and standard deviation of the Gaussian curve. The three parameters can be interpreted as follows. The area of the Gaussian relates to the electron density of the plasma in the voxel, the mean value corresponds to the mean temperature, and the standard deviation is an indication of how multi-thermal the plasma in the voxel is; the larger the width the more diverse the temperatures within the voxel. By computing the moments of the LDEM, we obtained the mean electron density and temperature in each voxel. Then, a 3D distribution of the mean electron and temperature of the global corona is constructed.

To obtain the coronal magnetic field we use a {potential-field source-surface \citep[PSFF, see][]{schrijver_2003}} model that extrapolates the magnetic field from {photospheric synoptic magnetograms obtained with the Michaelson Doppler Imager (MDI) on board SOHO}, used as boundary condition. {To perform the reconstruction we select the starting points at the center of each tomographic voxel, at uniformly spaced heights, every 0.02 $R_{\odot}$, within 1.025 and 1.225 $R_{\odot}$ and every 2\textdegree~in latitude and longitude. Thus, we obtain a magnetic field reconstruction that covers the whole volume that contains DEMT results. The source surface is set at 2.5 $R_{\odot}$, a height of $\approx1045$ Mm above the photosphere. Magnetic field lines are integrated using the PFSS Solarsoft package \citep{schrijver_2003} and associated to individual loops. In order to combine the magnetic field reconstruction with the 3D distribution of the plasma parameters, we use the DEMT results in the voxels crossed by each loop within the tomographic limits (1.025 to 1.225 $R_{\odot}$). Since the resolution of the field line is higher than the tomographic resolution, given a magnetic field line, we only keep one data point of the loop on each tomographic voxel that it crosses. We choose the middle point of the loop segment crossing the voxel. Then, we associate to that point the DEMT products (electron density and mean temperature) of the corresponding voxel. Thus, each magnetic loop is formed by as many data points as tomographic voxels it crosses.}

{In this work, we only analyze closed magnetic loops within the tomographic limits. As in previous works \citep{maccormack_2017,lloveras_2017}, these loops are separated in two legs, from the coronal base to the apex of the loop. Since we have the plasma parameters traced along each leg, we can apply one fit for each one, to ensure a good characterization of the thermal properties along the loops. In the case of the density, we observe a strong variation with height which is consistent in general with an exponential behaviour associated to a certain height-scale. We therefore use an exponential least-square fit whose quality is characterized by its coefficient of determination $r^2$\footnote{$r^2\equiv1-S_{res}/S_{tot}$, where $S_{res}$ is the sum of the squared residuals and $S_{tot}$ is the sum of data deviations from the mean.}. It is worth noting that we use an exponential fitting only for practical reasons, and that we are not assuming, in principle, that the reconstructed loops are necessarily in equilibrium.} For the temperature, the variation with height is much smoother, so we use a linear fit. In this case, we use the Theil-Sen estimator, which is more robust than the regular least-square fit, for its treatment of the relative weight of outliers data. To determine the success of the temperature fit, we consider which percentage of data along the loop falls within the uncertainty interval of the temperature calculation of the tomography ($\approx5-10\%$, see \citealt{lloveras_2017}). In this way, we obtain the variation of electron density and temperature along each loop, as well as the height scales and gradients needed to compute the energy flux, along the loops. In the next section we present the energy balance model used to compute this energy flux.\\

\subsection{Energy balance model}
\label{model}

{In order to obtain a rough estimation of the heating injected into the loops}, we assume an energy balance situation in which all gains are compensated by losses. In this way, for a coronal magnetic flux tube, the coronal heating power ($E_h$) is locally balanced by the radiative ($E_r$) and the thermal conduction ($E_c$) losses \citep{aschwanden_2004}. Thus, if $s$ is the position along the flux tube, we have:

\begin{equation}\label{Balance}
E_h(s) = E_r(s)+ E_c(s),
\end{equation}

\noindent
where the quantities are in units of $[{\rm erg\,sec^{-1}\,cm^{-3}}]$. To compute the thermal conduction power, in this particular plasma regime we use the Spitzer model \citep{spitzer_1962}, in which the thermal conduction is associated to the divergence of the conductive loss function:

\begin{equation}\label{Fc}
F_c(s)=-\kappa_0\,{T(s)}^{5/2}\,\frac{dT}{ds}(s). 
\end{equation}

\noindent
where $\kappa_0$ is the Spitzer thermal conductivity $\kappa_0 = 9.2 \times 10^{-7}  {\rm erg\,sec^{-1}\,K^{-7/2}}$.

Since thermal conduction is strongly confined to the magnetic field direction, the divergence is simply the derivative along the position $s$,

\begin{equation}\label{Ec}
E_c(s)=\frac{1}{A(s)}\,\frac{d}{ds}\left[A(s)\,F_c(s)\right].
\end{equation}

\noindent
where $A(s)$ represent the loop area along the position $s$.

Assuming a quasi-isothermal plasma approximation in the quiet-Sun corona, the radiative power can be expressed as plasma density to the second power multiplied by a radiative loss function, $\Lambda(T)$, that accounts for the temperature dependence of the radiated emission. To calculate the radiative loss function we use the atomic database and the plasma emission model from CHIANTI \citep{delzanna_2015}. We obtain the electron density from the DEMT results. We then compute the radiative power along the flux tube as:

\begin{equation}\label{Er}
E_r = \int\,\Lambda(T)\,dN_e^2(T).
\end{equation}

Integrating each of the three power quantities, $E_h, E_r, E_c$, over the volume of the magnetic flux tube, and dividing the result by its basal area, a loop-integrated version of the energy balance is obtained,

\begin{equation}\label{FluxBalance}
\phi_h = \phi_r + \phi_c.
\end{equation}

\noindent
where the three resulting loop-integrated quantities $\phi$ have units of energy flux $[\rm{erg\,sec^{-1}\,cm^{-2}}]$, and the equation holds now for each individual field line (as opposed to flux tube). 

{Due to the null divergence condition of the magnetic field, it can be integrated along the magnetic flux tube to obtain the following relation with the flux tube area: $A(s)\,B(s)=A_0\,B_0=A_L\,B_L$, where $B_0$ and $B_L$ are the values of the magnetic field at the footpoints of the loop in the coronal base. Therefore, the radiative and conductive terms can be rewritten as}

\begin{eqnarray}
\phi_r &=& \left(\frac{B_0\,B_L}{B_0+B_L}\right)\,\int_{0}^L ds\,\frac{E_r(s)}{B(s)}\label{phir2},\label{phir2}\\
\phi_c &=& \frac{B_0\,F_{c,L}-B_L\,F_{c,0}}{B_0+B_L} \label{phic}.\label{phic2}
\end{eqnarray}

For a fully detailed description of the energy balance model, we refer the reader to \citet{maccormack_2017}.\\

\subsection{Data}
\label{Data}

In this work we use EUV images from the \emph{Extreme Ultraviolet Imager} (EUVI, \citet{wuelser_2004}) on board the \emph{Solar Terrestrial Relations Observatory} (STEREO) mission \citep{kaiser_2008}. We reconstruct the solar corona along Carrington rotation (CR) 2082 using three different wavelength bands of the telescope: 171~\AA, 193~\AA, and 284~\AA, which have a maximum temperature sensitivity in the range $1.0-2.15$ MK (see \citet{nuevo_2015}). This rotation started on 5 April 2009 and finished on 3 May 2009 during the minimum between Solar Cycles 23 and 24. 

DEMT provides an average description of the coronal state during the data-acquisition time (half a synodic solar rotation period), and cannot temporally resolve coronal dynamics. If ARs are observed during the analyzed rotation, the DEMT results may produce artifacts such as negative density values on the voxels covering active events. We call these voxels zero-density artifacts (ZDAs) and they are discarded from the analysis. No significant active regions (ARs) were observed during Carrington rotation 2082.\\

\subsection{Method}
\label{Method}

In this work we focus on closed magnetic loops whose apexes are within the limits of the tomographic box ($\approx14$ Mm to $\approx150$ Mm). Each reconstructed loop must satisfy certain criteria to be considered viable for analysis. These criteria were already used in previous works \citep{maccormack_2017,lloveras_2017}. First, separating the loops in two legs, each leg must count with at least five viable voxels (without ZDAs) to ensure a good determination of the parameters, otherwise the loop is discarded. Secondly, the {coefficient of determination} for the density fit, $r^2$, must exceed 0.75 on each leg. For the temperature, it is required that at least {$50\%$} of the data along each leg be contained within the uncertainty interval of the tomographic temperature computation. We consider that following these criteria a reliable set of closed coronal loops is obtained. {We start with $\approx55000$ loops and, after applying these criteria, we keep $\approx30000$ loops with reliable data.}

In previous works, it was found a dependence of the temperature on the latitude where the loop was located \citep{nuevo_2015,lloveras_2017}. They observed that structures at latitudes close to the equator ($-20,20$\textdegree) had cooler temperatures than loops at middle latitudes ($|20,60|$\textdegree). Considering this, we divide our loop population into three regions: south latitudes $(-20,-60)$\textdegree, equator latitudes $ (-20,20)$\textdegree  and north latitudes $(20,60)$\textdegree. Here, we consider that a loop corresponds to a certain latitude range if its two footpoints lie on that range. Because of this selection criteria, the loops in our set tend to be of medium and short length (less than 800 Mm) and are located within the coronal streamer. Loops that cross the equator and surround the streamer are not included, and neither are loops open to the interplanetary medium (which we call ``open loops'').

Once we have reconstructed and selected the loops in each region, we compute loop-average properties for each of them. Each loop is characterized by its length, loop-average density, temperature, and pressure along their length (starting from $\approx14$ Mm above the photosphere), loop-average magnetic field (starting from the photosphere) and loop-integrated energy flux quantities obtained with the energy balance model (see Section \ref{model}) at $\approx14$ Mm of height. {We average the magnetic field starting from the photosphere instead of a height of $\approx14$ Mm, in order to obtain a more substantial variation of the mean magnetic field for different loops in the set, since the field varies much less along coronal heights.}  

Since we have large samples of loops (of the order of 5000 on each region), to simplify the analysis we proceed to separate them in small length bins, obtaining loop-average values of the parameters of interest within each bin. In each region, there is a distribution of loop lengths ranging from $\approx100$ Mm to $\approx800$ Mm. We divide the loop length range in 35 non-uniform bins, set so that each one contains the same number of loops {to have the same statistical noise}. We have verified that our results do not change significantly if we vary the number of bins. On each bin we compute the median of the parameters of interest and an estimated error. The error is obtained from the median of the differences between individual loop-averages and the median of the bin. We consider that following this procedure we obtain more robust fits for the analysis of the relations between loop parameters. We present our results in the next section.\\

\section{Results}
\label{Results}

\subsection{Loop length distribution}\label{L}

\begin{figure*}[ht]
\begin{center}
\includegraphics[width=0.32\textwidth]{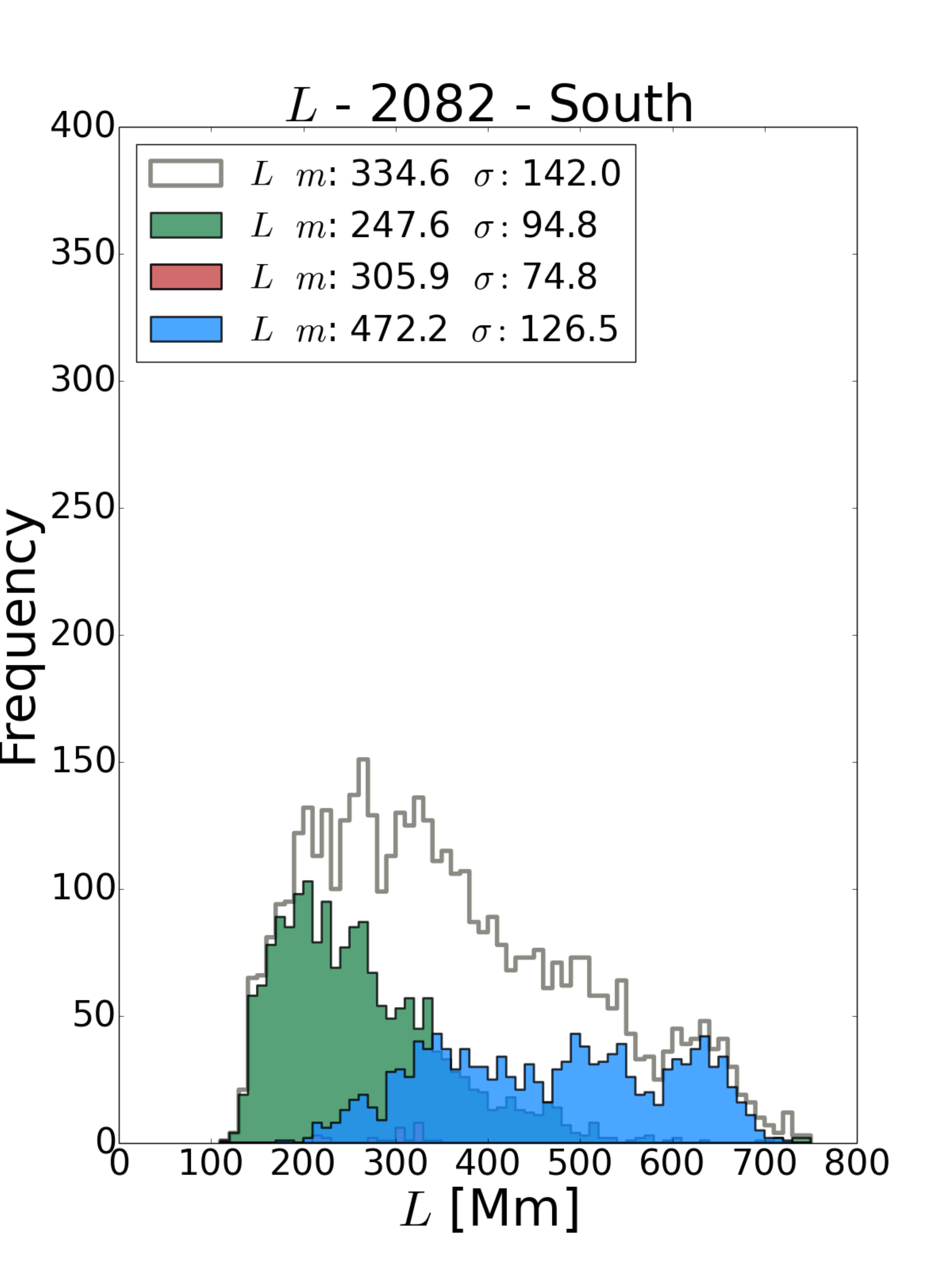}
\includegraphics[width=0.32\textwidth]{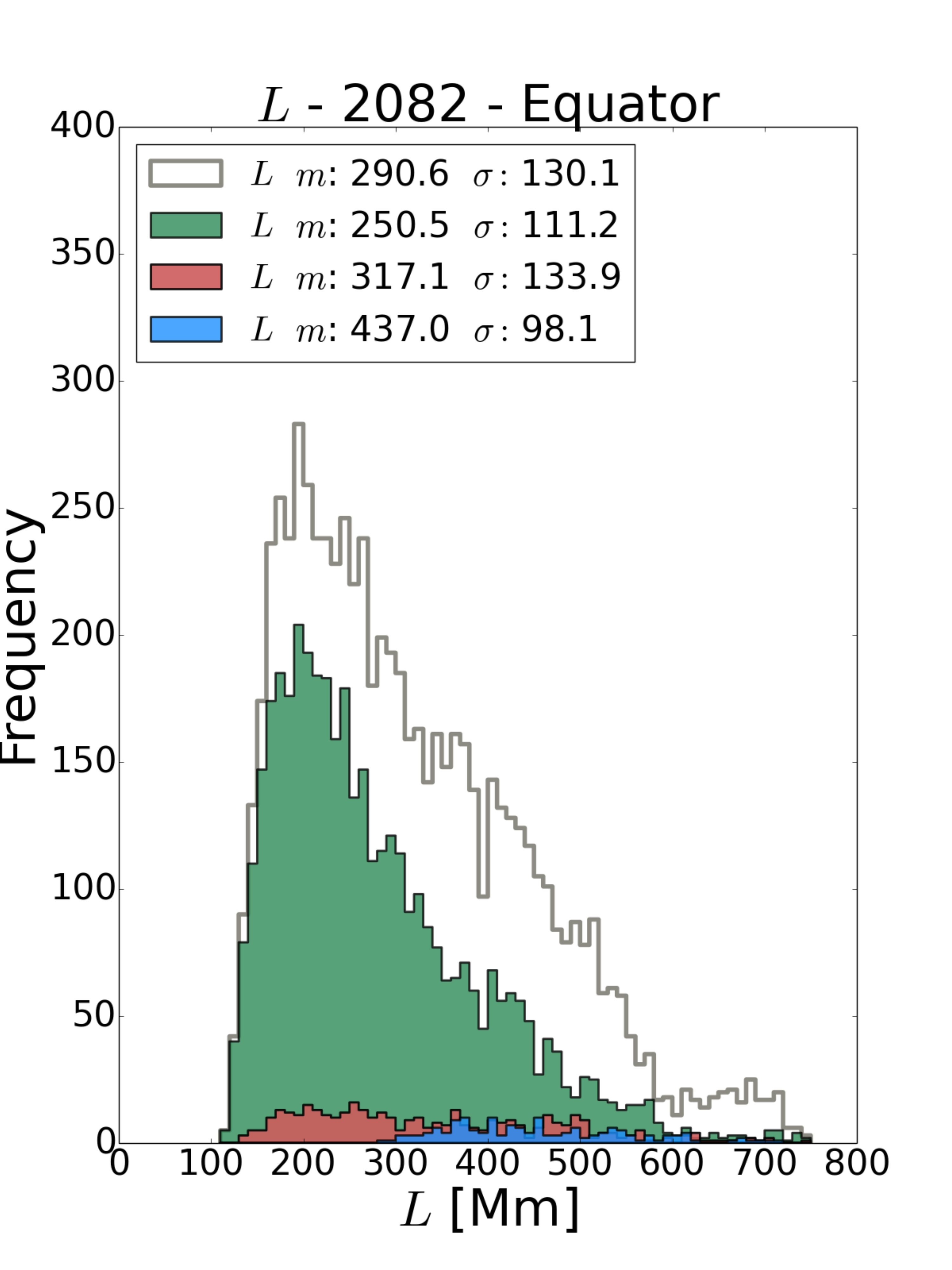}
\includegraphics[width=0.32\textwidth]{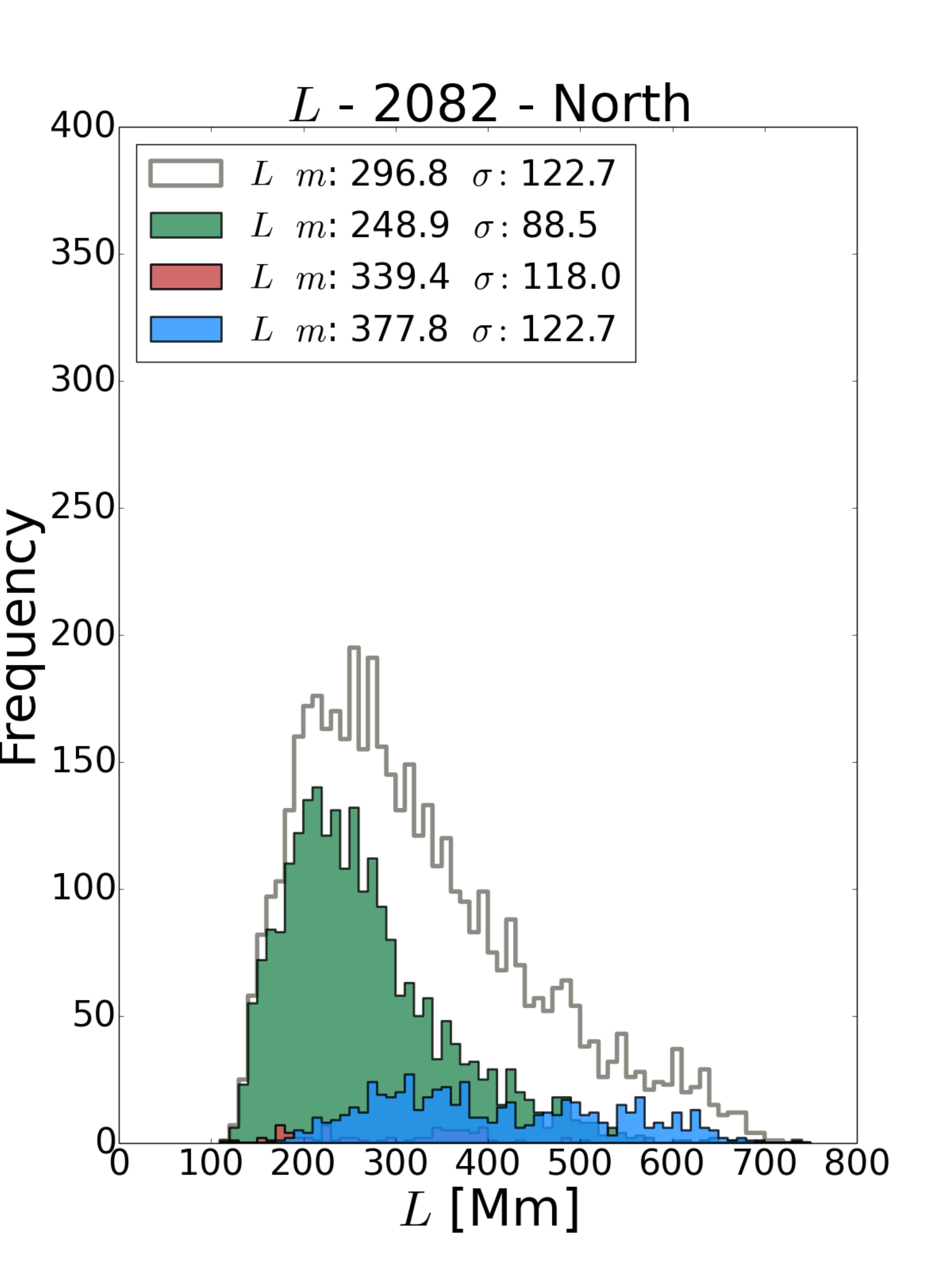}
\includegraphics[width=0.80\textwidth]{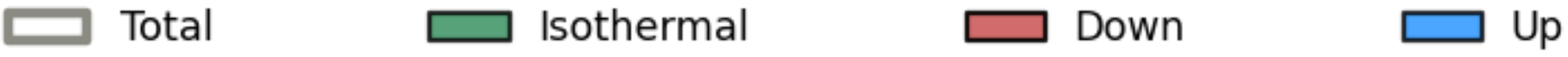}
\caption{Frequency histograms of loop length, $L$, for the different latitude regions: south $(-20,-60)$\textdegree (left panel), equator $(-20,20)$\textdegree~(middle panel) and north $(20,60)$\textdegree~(right panel). The gray line indicates the total number of loops with both footpoints in the indicated latitude. Green, blue, and red histograms correspond to isothermal, up, and down loops respectively. In all cases, the median value, $m$, and the standard deviation, $\sigma$, are indicated.}
\label{HistL}
\end{center}
\end{figure*}

Fig. \ref{HistL} shows the distribution of loop lengths $L$ on each latitude region. The gray line represents the total population of analyzed loops. As already mentioned, the loops in the selected set are relatively short and are located within the streamer. The median length is $\approx300$ Mm at all latitudes. However, it can be noticed that there is a substantial population of shorter loops around a length of $\approx150$ Mm in the equatorial region.

In previous works \citep{huang_2012,nuevo_2013}, two classes of loops have been identified according to the temperature variation along {the coronal part} of their length. Magnetic structures whose temperature increase or decrease with height were classified as up or down loops, respectively. In an up loop, the temperature is higher at the apex than at the coronal base, and the opposite in a down loop. In those articles, the authors found that down loops are mainly located at the equatorial region. \citet{serio_1981} were the first to analyze these structures, {finding a temperature inversion along the coronal part of the loops}. They also found that for down loops to be thermodynamically stable their length should be less than 3 times their density scale height. This conclusion is consistent with our results, since down loops tend to be shorter while up loops have more evenly distributed lengths.

In this work, we classify up and down loops according to the following criteria. We consider that a loop is up if its temperature gradient is positive in both legs and the temperature variation between the foot and the apex is greater than the median of the standard deviation of the temperature distributions (the LDEMs) in the voxels that are crossed by the leg. If both legs meet the criteria, the loop is called up. In this way, it is ensured that the temperature variation is substantial, because it must be larger than the characteristic standard deviation of the plasma thermal distribution. This variation is also much larger than the typical error of the tomographic method \citep{lloveras_2017}. Similarly, for a loop to be called down, the temperature gradient must be negative in both legs of the loop and the absolute value of the variation between the foot and the apex must be greater than the median of the width of temperature distributions. Finally, a loop is called isothermal if the temperature variation between the foot and the apex is less than the median of the thermal widths. These criteria must also be met in both legs of the loop. In all cases in which one leg meets one of the criteria, and the other one another, the loop is not labeled with any of these definitions.

It is worth to clarify that these loops are not properly isothermal, since the existence of a width in the thermal distribution of the voxels, and therefore the loops that cross them, are indicative of a multi-thermal plasma. However, it can be assured that their temperatures do not vary beyond the range determined by the median of the thermal widths. In the same sense, according to our criteria, up and down loops are considered such, if their temperature variations along the loop length is larger than  the range determined by the loop-average thermal width, i.e., the temperature variation is beyond the normal multi-thermality of the plasma. To class loops as isothermal, up or down, previous works \citep{lloveras_2017}, considered variations larger or smaller than the range defined by the error interval of the tomographic procedure. The criteria applied here are more strict in terms of relevant variations of temperature along the loops. 

In Fig. \ref{HistL} we show histogram distributions of isothermal loops (green histogram), up loops (blue histogram) and down loops (red histogram), separated by heliographic latitude region. We observe that in each latitude there is a large population of isothermal loops that follow the behaviour of the full set. We note that the largest population of down loops is found around equatorial latitudes. These loops have dominantly short lengths between $\approx100$ Mm and $\approx400$ Mm, as expected. up loops, on the other hand, dominate the south and north latitudes, with lengths between $\approx200$ Mm and $\approx700$ Mm.

In order to analyze the behavior of the loop-average parameters as a function of length, for each latitude we separate the loop lengths in 35 bins, as described in Section~\ref{Method}. In the following sections we present the results for the different loop parameters.\\

\subsection{Density and pressure}\label{NP}

\begin{figure*}[ht]
\begin{center}
\includegraphics[width=0.32\textwidth]{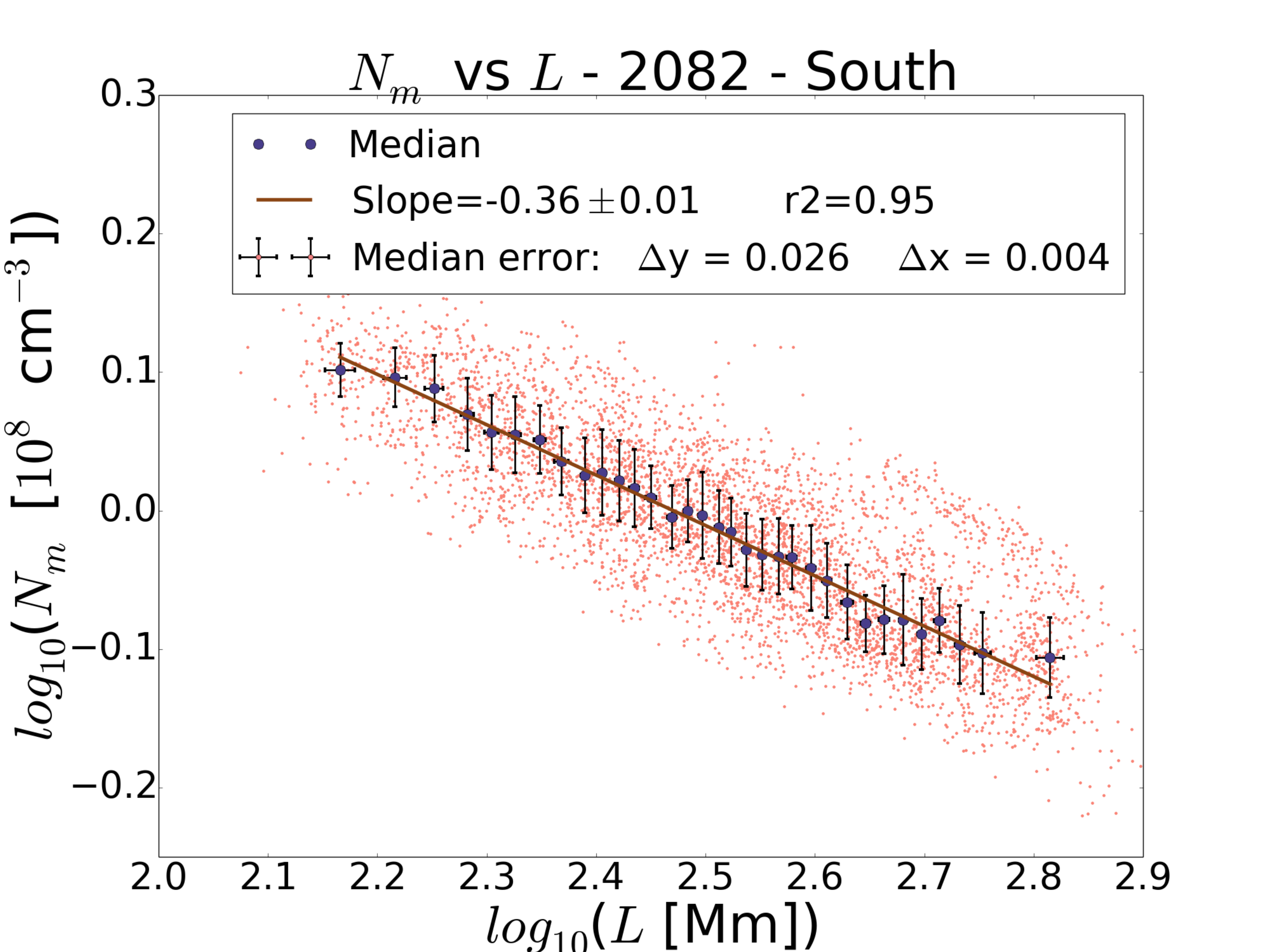}
\includegraphics[width=0.32\textwidth]{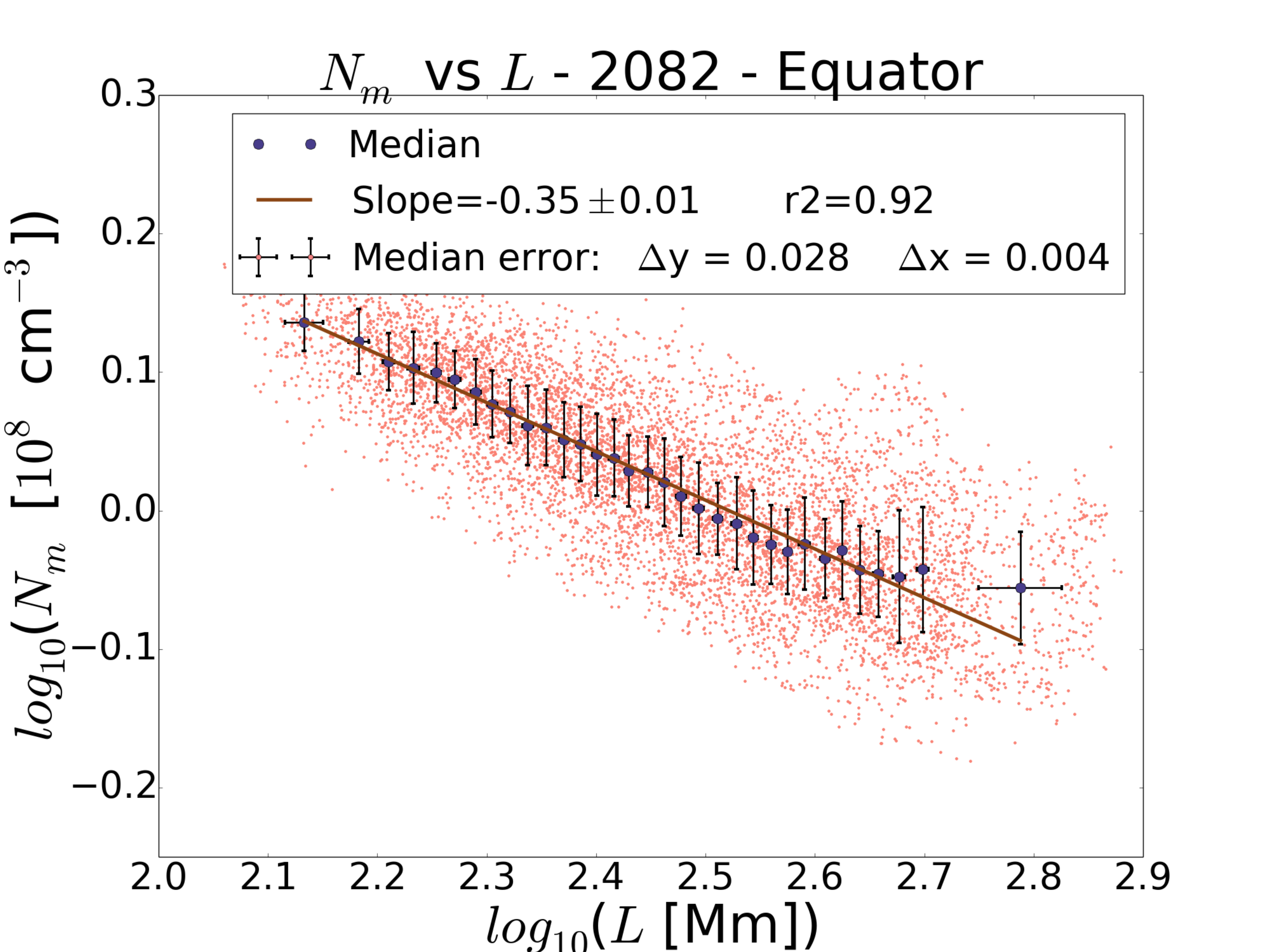}
\includegraphics[width=0.32\textwidth]{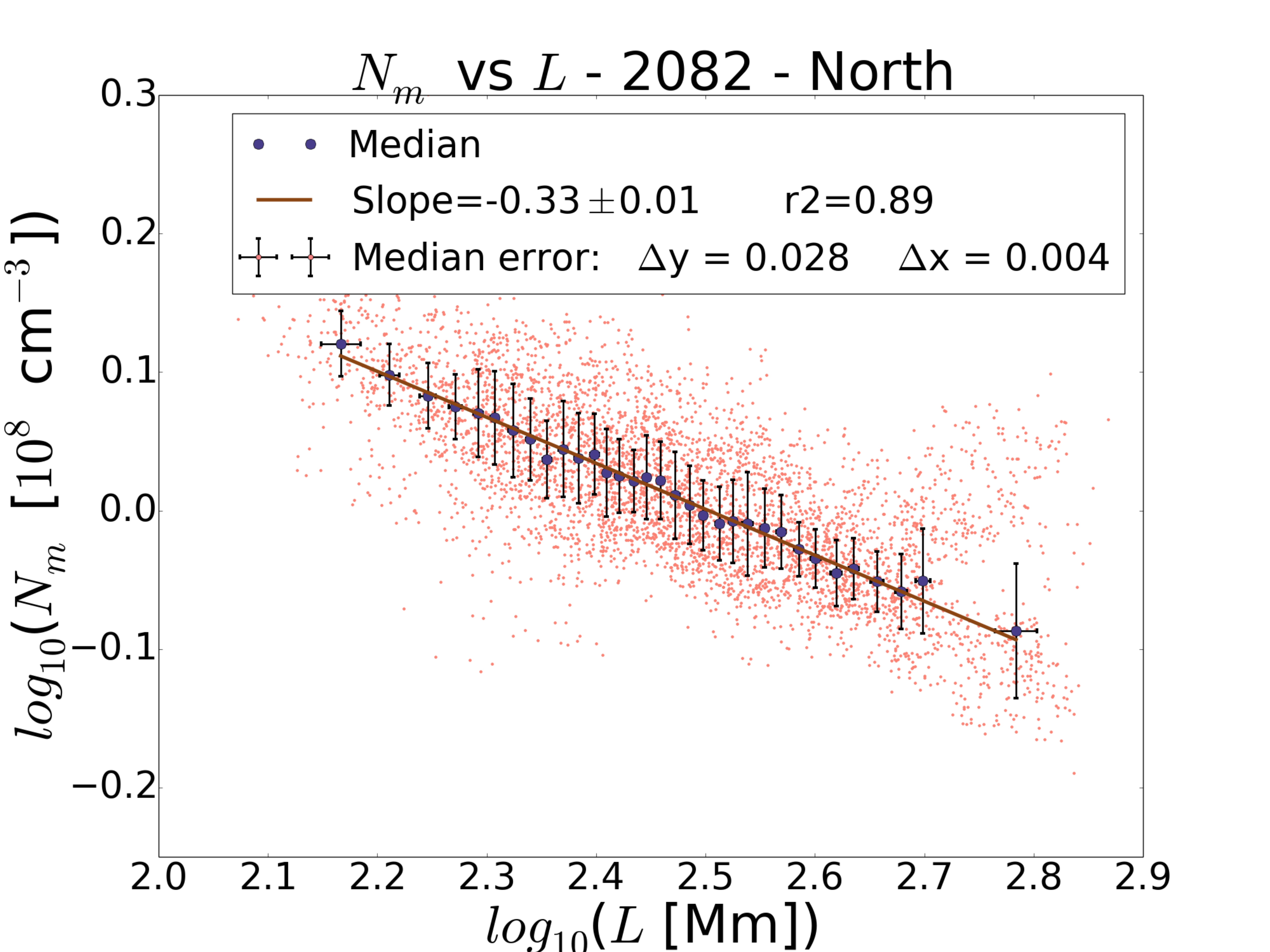}
\caption{Scatter plots of loop-average density $(N_m)$ vs. length $(L)$ in each latitude: north (upper panel), equator (middle panel) and south (bottom panel). Small pink dots represent the loop-average density starting from a height of $\approx14$ Mm. The bold-dark bullets represent the median of the loop-average density within loop length bins as described in Section \ref{NP}. Error bars represent the median of the distances between individual loop-averages and the median value of each bin. The continuous line is a linear fit of the median values of the bins. The fit results are accompanied by their corresponding coefficient of determination $r^2$ and the median error in the panel insets.}
\label{NvsL}
\end{center}
\end{figure*}

Fig. \ref{NvsL} shows logarithmic plots of density versus length. Each pink point in the data cloud corresponds to the loop-average density $N_m$ of each analyzed loop and the bold-dark bullets correspond to the median of the loop-average densities per bin as a function of loop length median per bin. The accompanying error bars correspond to the standard deviation of the density distribution within each bin. A decreasing behavior is observed as expected, since the plasma is denser near the coronal base, so longer loops tend to have lower loop-average densities than short ones. We also note that, for the three latitude regions, the slopes found are very similar, $\approx -0.35$, implying a scale law of the type $N_m\approx L^{-0.35}$. In all cases the coefficient of determination, $r^2$, is equal or higher than 0.85.

Due to the validity of the ideal gas approximation in the quasi-isothermal corona, the behavior of the loop-average pressure is very similar to that of the density. We obtain in this case similar slopes, $\approx-0.3$ in the three latitudes, all of them with a coefficient of determination, $r^2$, higher than 0.85.\\ 

\subsection{Temperature and magnetic field}\label{TB}

\begin{figure*}[ht]
\begin{center}
\includegraphics[width=0.32\textwidth]{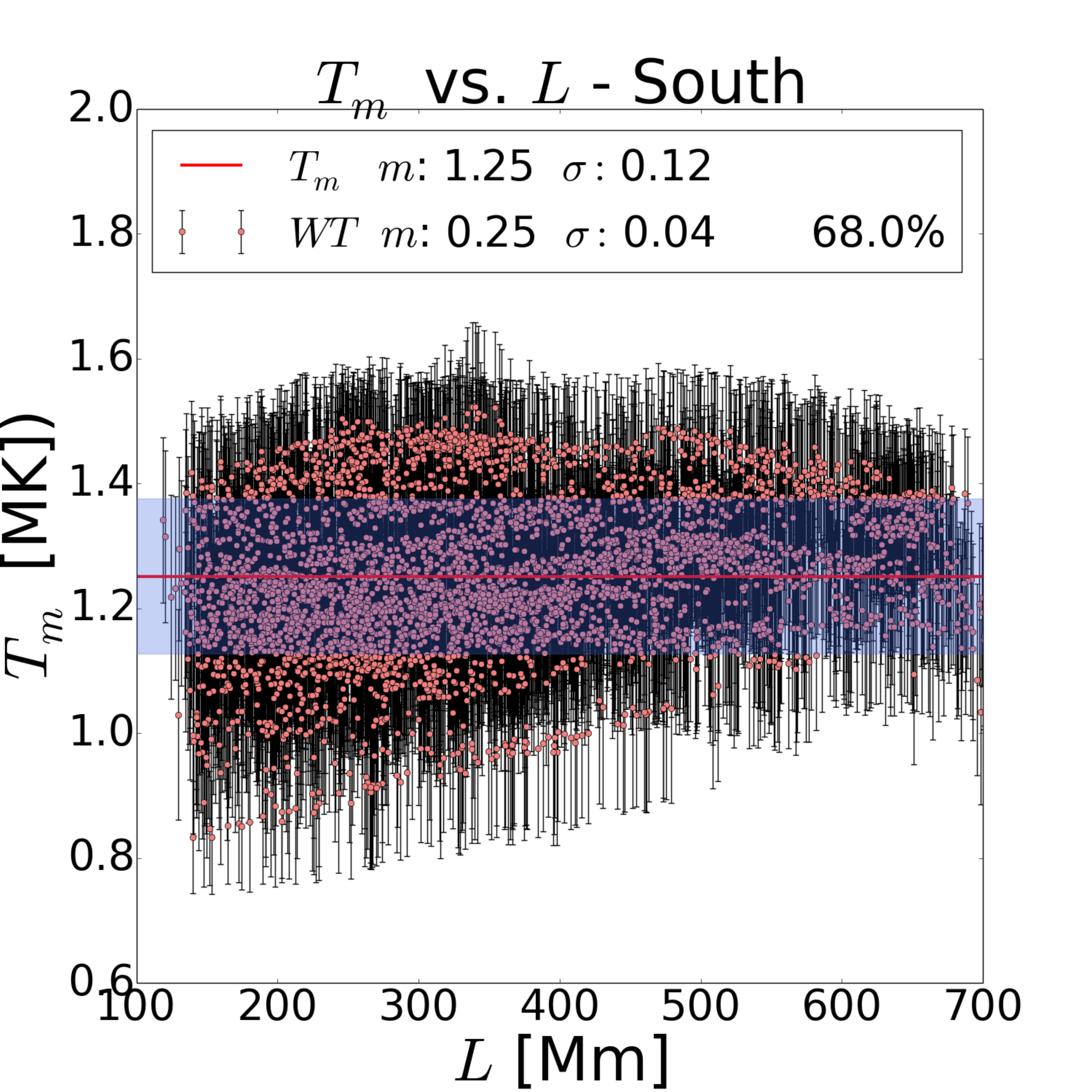}
\includegraphics[width=0.32\textwidth]{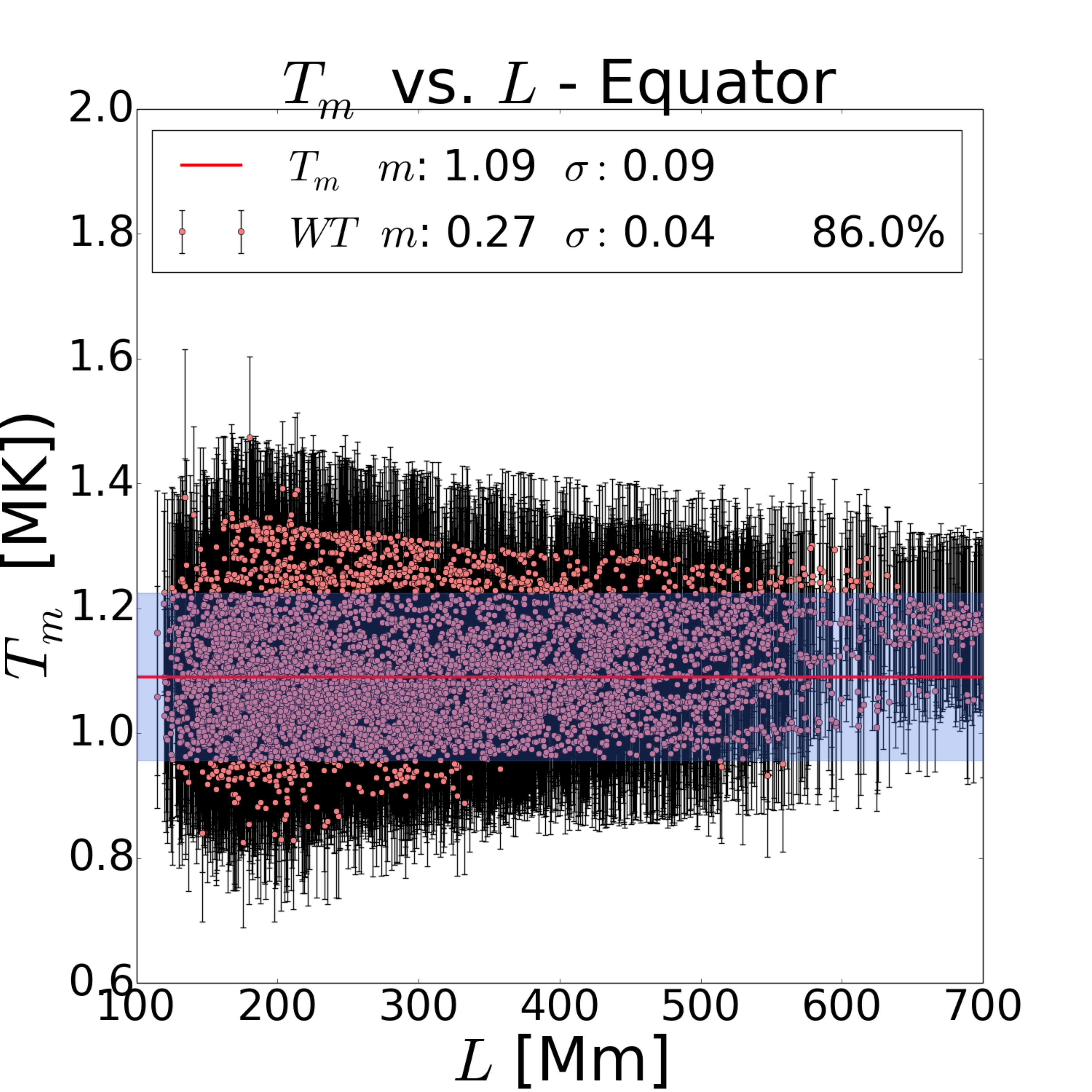}
\includegraphics[width=0.32\textwidth]{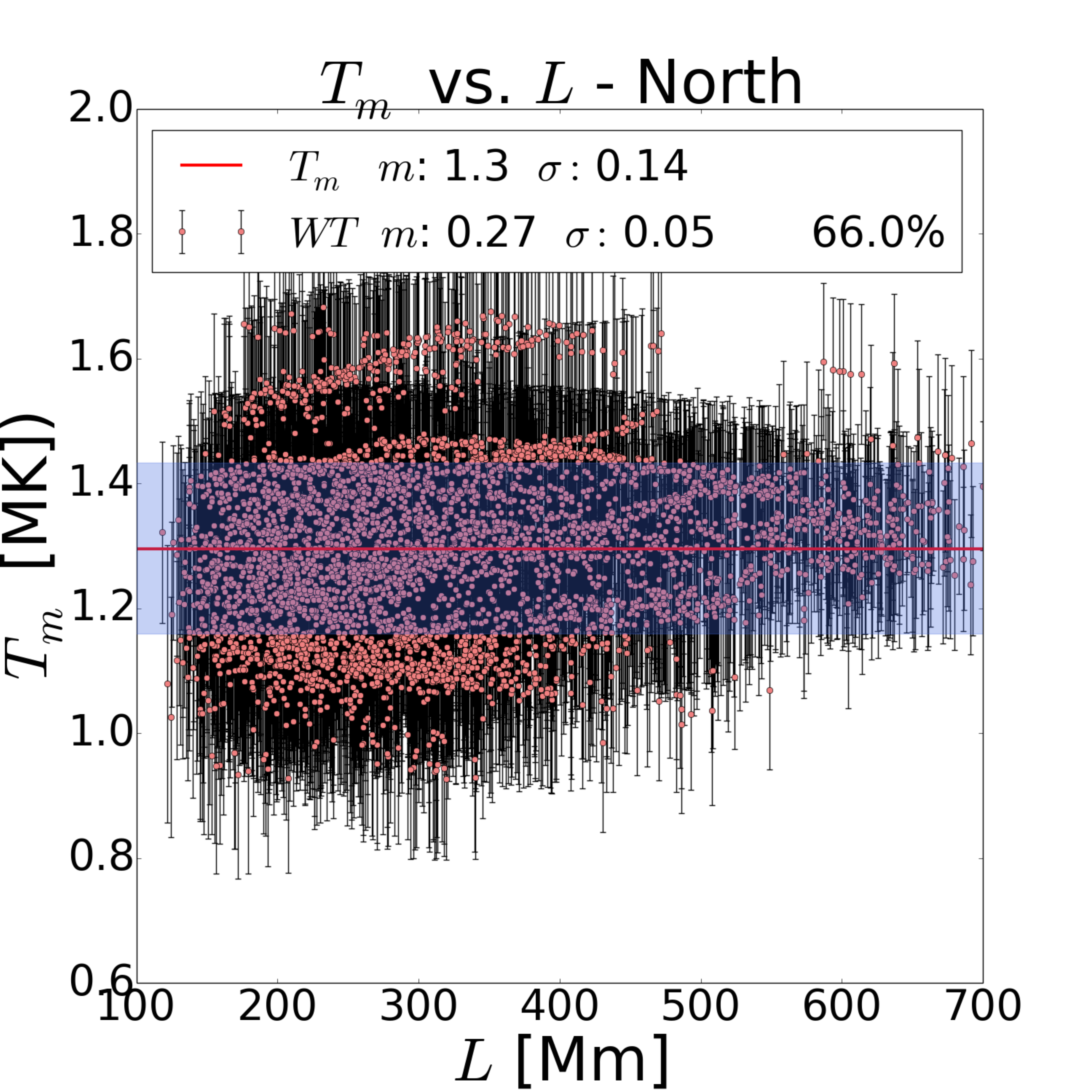}
\caption{Scatter plots of loop-average temperature $T_m$ vs. length $L$ for three latitudes: south (left panel), equator (middle panel) and north (right panel). Small dots represent the loop-average temperature in each latitude from $~14$ Mm of height and error bars indicate the loop-average temperature distribution width $WT_m$. Continuous red lines are the median of the loop-average temperatures of all loops and the blue shaded area represent the median of the temperature distribution width of all loops. We show the percentage of loops that are in the blue area, and that are considered ``average loops''.}
\label{TvsL}
\end{center}
\end{figure*}

Studying the temperature as a function of length, we do not find the same clear dependence as for density and pressure. However, analyzing each latitude region separately, we find that the loop-average temperature $T_m$ in equatorial latitudes is lower than in the north and south latitudes. This is consistent with previous solar minima studies showing that down loops are mainly located at low latitudes \citep{nuevo_2013,lloveras_2017}. Fig. \ref{TvsL} shows the loop-average temperature of each loop as a function of loop length. The accompanying error bars are the loop-average width of the temperature distributions obtained from the LDEM widths. The red solid line is the median of all loop-average temperatures in the latitude region, and the width of the blue shaded area around the temperature median corresponds to the median of all the loop-average thermal widths $WT_m$. In the upper left corner of the panels, we indicate the percentage of loops whose loop-average temperatures lie within the blue shaded area. Therefore, we associate the blue area with the typical temperature and thermal width of the loops in each latitude region. It can be seen that the loop-average temperature in equatorial latitudes is $\sim15\%$ lower than in middle-latitudes, as it has been observed in previous works \citep{maccormack_2017}.

\begin{figure*}[ht]
\begin{center}
\includegraphics[width=0.32\textwidth]{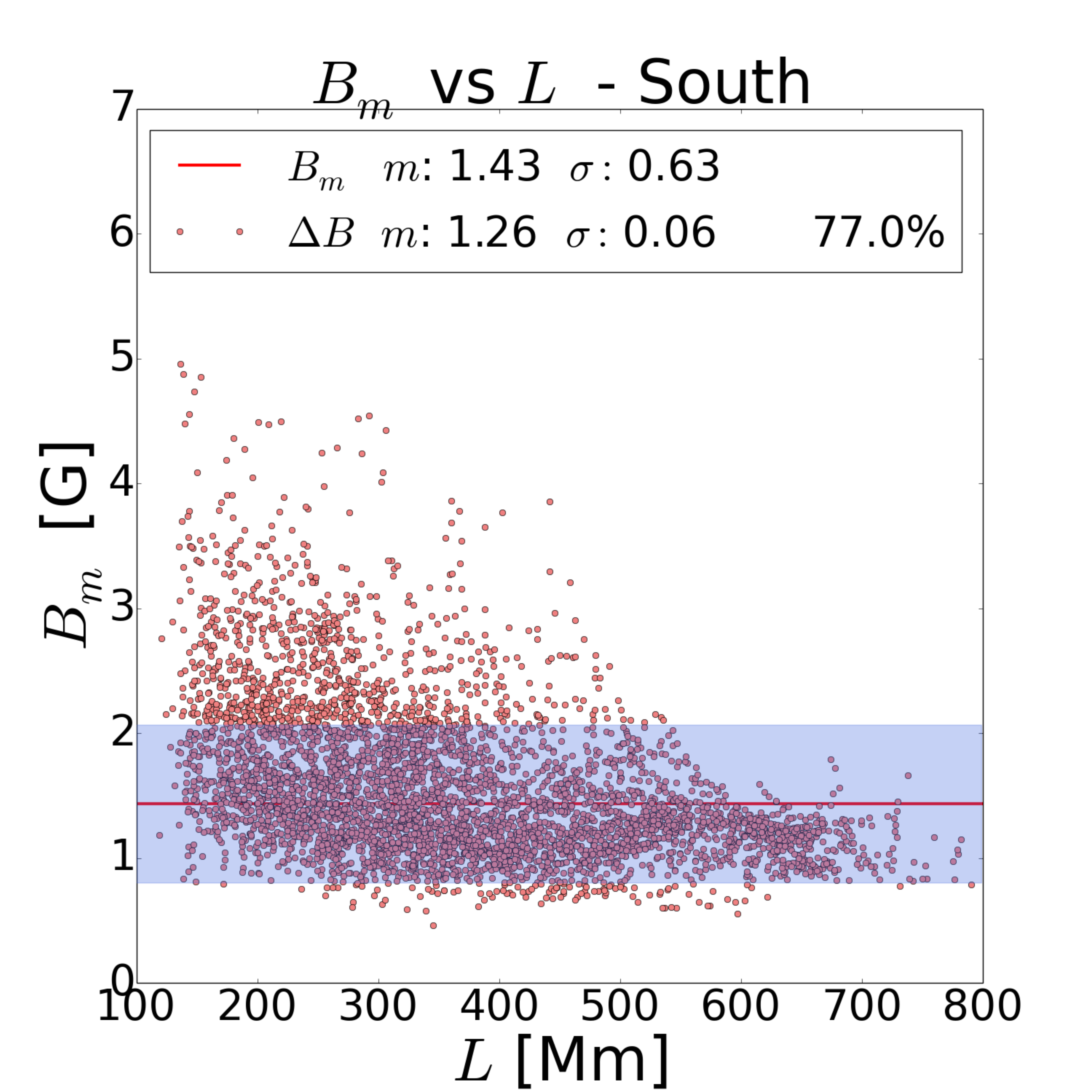}
\includegraphics[width=0.32\textwidth]{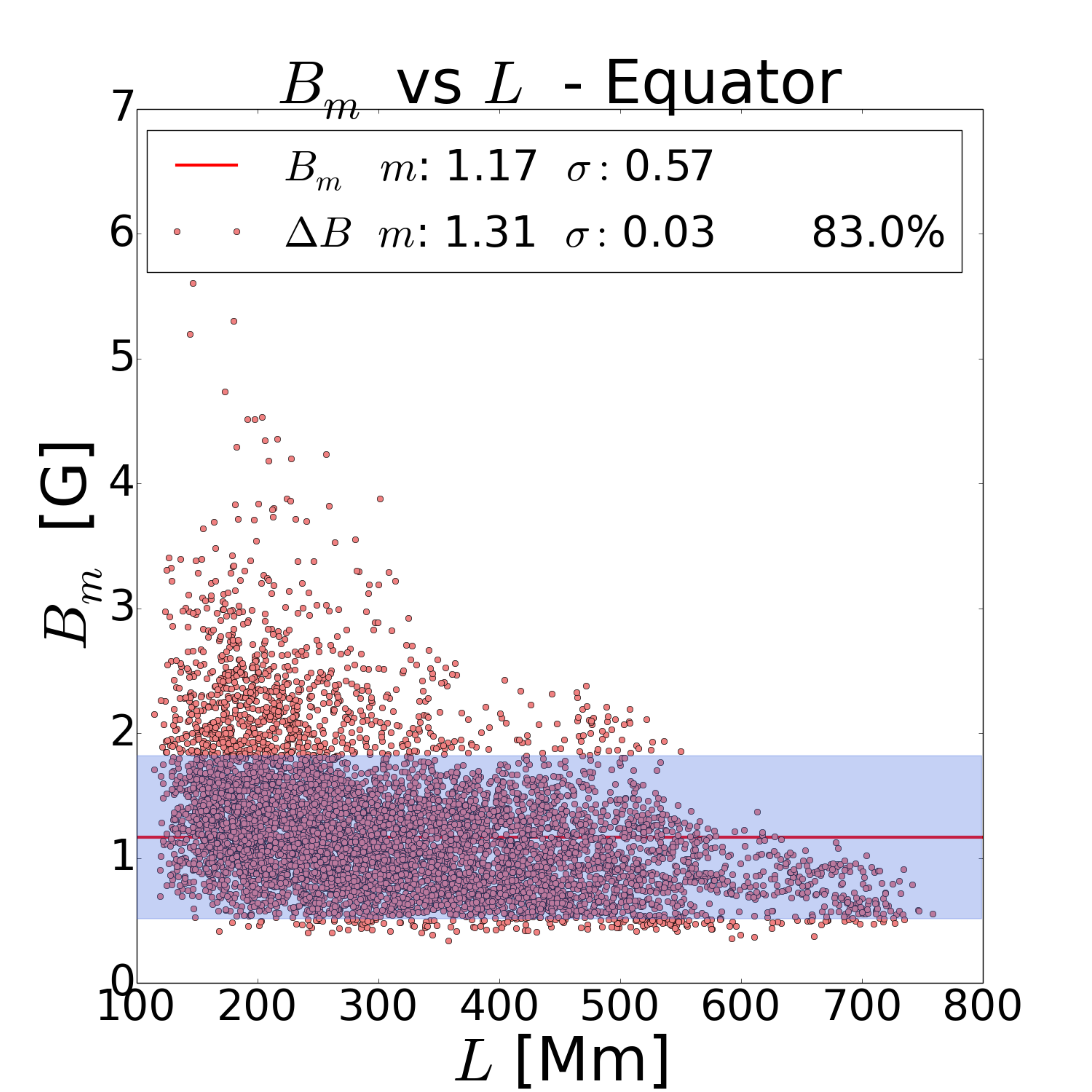}
\includegraphics[width=0.32\textwidth]{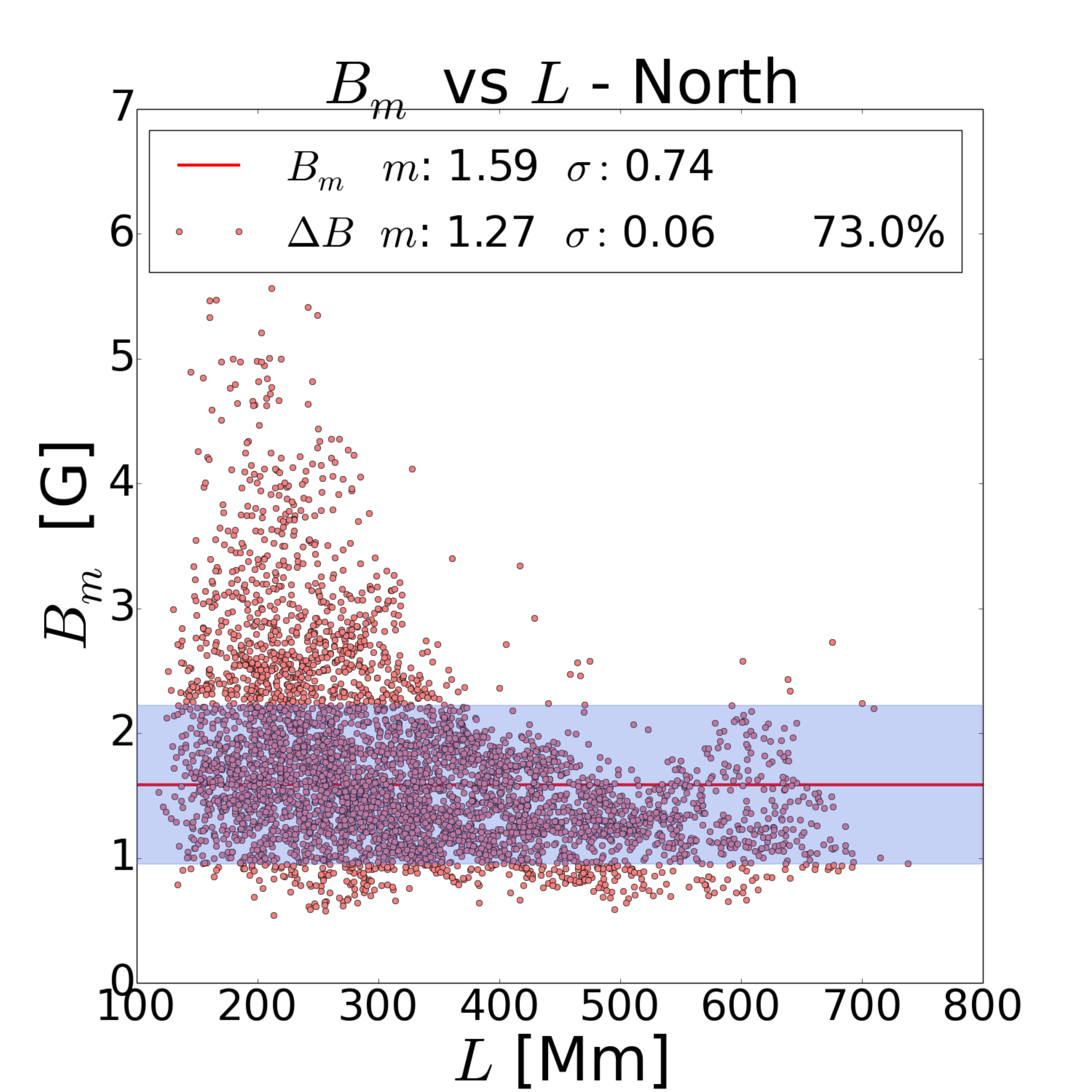}
\caption{Scatter plots of loop-average magnetic strength $B_m$ vs. length $L$ for three latitudes: south (left panel), equator (middle panel) and North (right panel). Small dots represent loop-average magnetic field starting from the photosphere. Continuous red lines correspond to the median of the loop-average magnetic field of all loops and the blue shaded areas represent the median of the differences between the median value of magnetic field of all loops and the global median. We show the percentage of loops that are in the blue shaded area, implying that they should be considered ``average loops''.}
\label{BvsL}
\end{center}
\end{figure*}

In Fig. \ref{BvsL} we show the loop-average magnetic field $B_m$ as a function of loop length. The solid lines represent the median of the magnetic field in the corresponding latitude region, and the blue shaded area corresponds to the median of the differences between each individual loop value and the magnetic field median. It can be seen that the loop-average magnetic field has a similar behavior to the temperature. We find median values that are $\approx20\%$ smaller in equatorial latitudes that in south and north latitudes. {Also, a decreasing behavior is observed as a function of length, which indicate that the longer the loop, the smaller the loop-average magnetic field. Looking for a scaling law of the type $B_m \sim L^{-r}$, we obtained $r$ values in the range $\approx[0.15,0.55]$ in the three latitude ranges. However, the fit quality parameters vary between 0.45 and 0.8 in the three latitude regions.}

\begin{figure*}[ht]
\begin{center}
\includegraphics[width=0.70\textwidth]{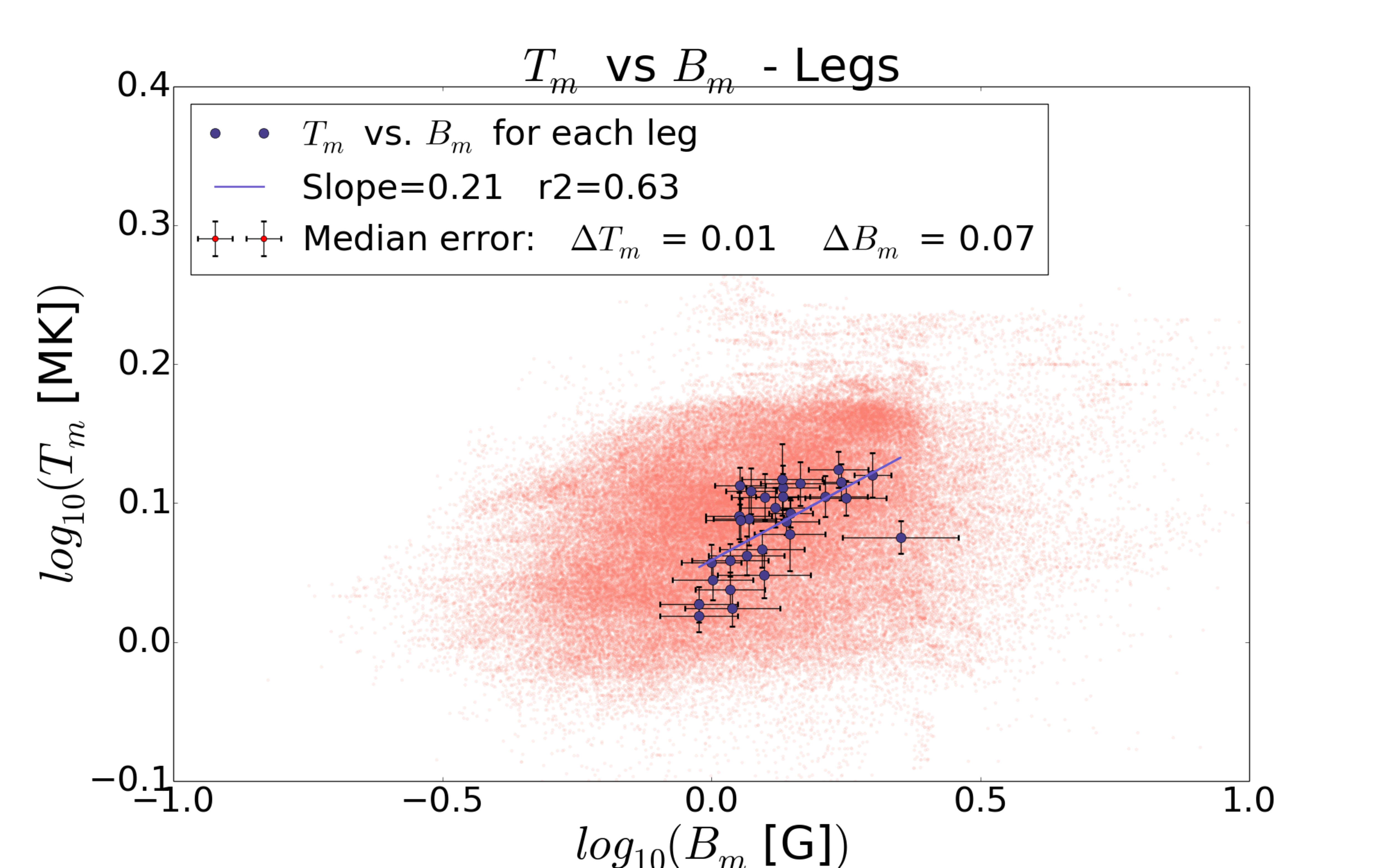}
\caption{Red dots correspond to the plot of the $T_m$ vs. $B_m$ for the legs of each reconstructed loop. Bold-dark bullets represent the median $(B_m,T_m)$ value within bins of 4\textdegree of latitude. Error bars correspond to the median of the differences between the each individual value and the median value of each latitude interval. The continuous line is the linear fit obtained with the median values of each interval. In the upper inset we show the fit results with its corresponding coefficient of determination $r^2$ and the median error for the temperature and the magnetic field.}
\label{TvsB}
\end{center}
\end{figure*}

Although we did not find a clear relation between temperature and magnetic field of the loops, the similar dependence on latitude found in both cases motivated us to perform the following analysis. We divide both datasets in bins of 4\textdegree of latitude and we compute the median of $T_m$ and $B_m$ in each bin. Since both loop legs do not necessarily lie in the same bin, we do this analysis by leg and not by loop, so we include on each latitude bin all the loop legs whose footpoint lies in that latitude. In Fig. \ref{TvsB} we show a logarithmic plot of the median temperature as a function of the median magnetic field for each latitude bin. A linear fit provides a slope of $\approx0.2$, suggesting a scale-law of the type $T_m \approx B_m^{0.2}$.

\subsection{Temperature distribution width}\label{WT}

\begin{figure*}[ht]
\begin{center}
\includegraphics[width=0.45\textwidth]{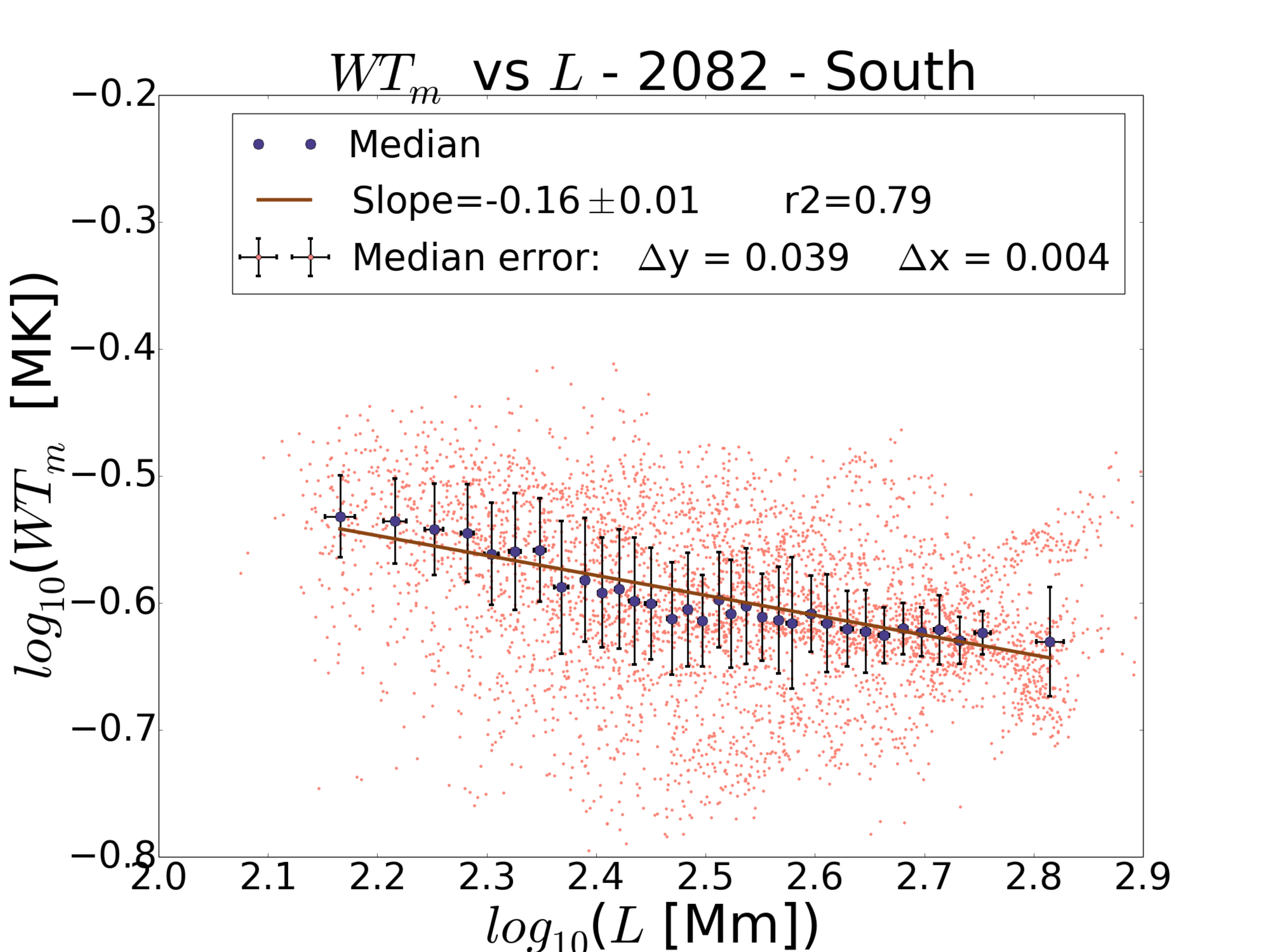}
\includegraphics[width=0.30\textwidth]{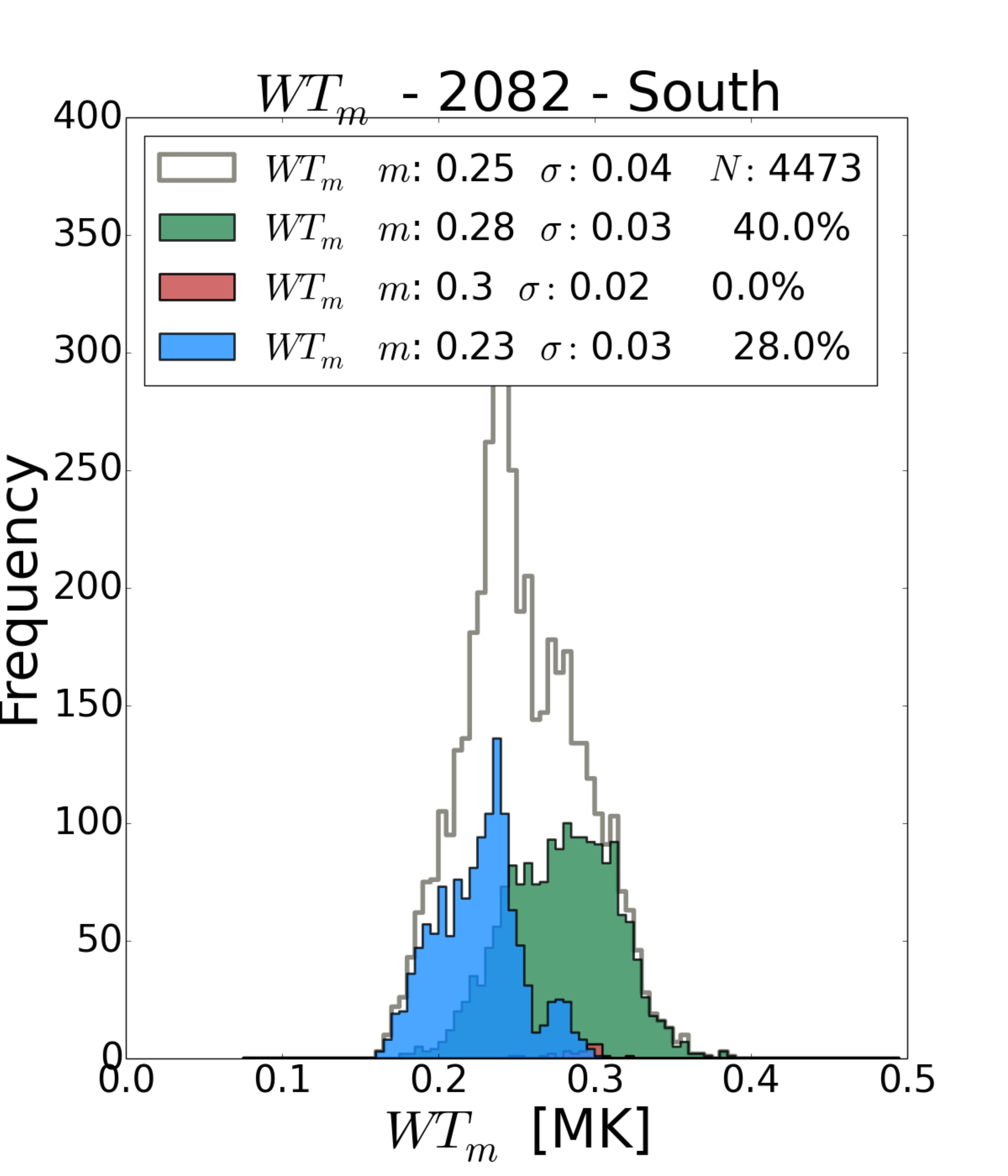}
\includegraphics[width=0.45\textwidth]{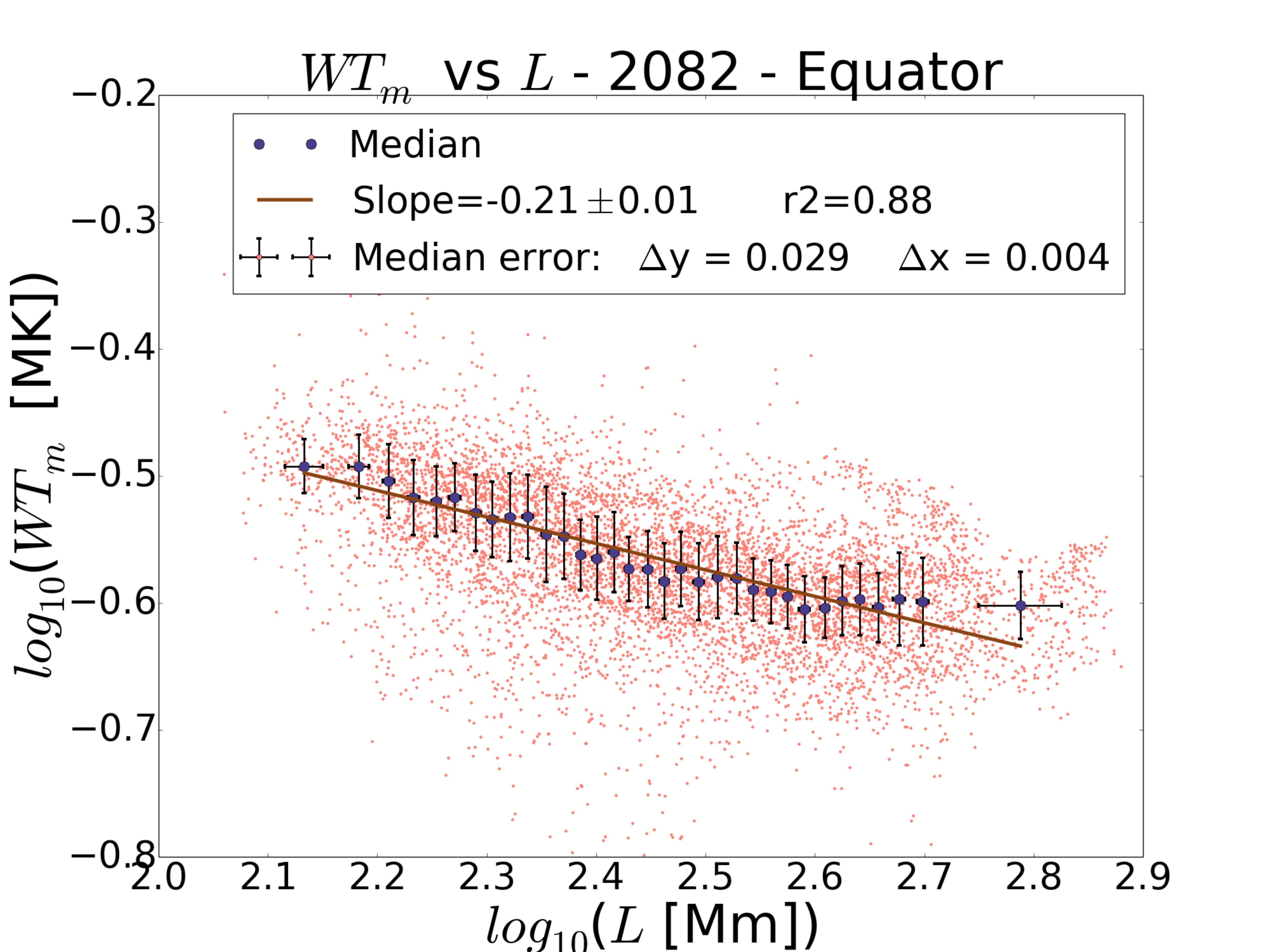}
\includegraphics[width=0.30\textwidth]{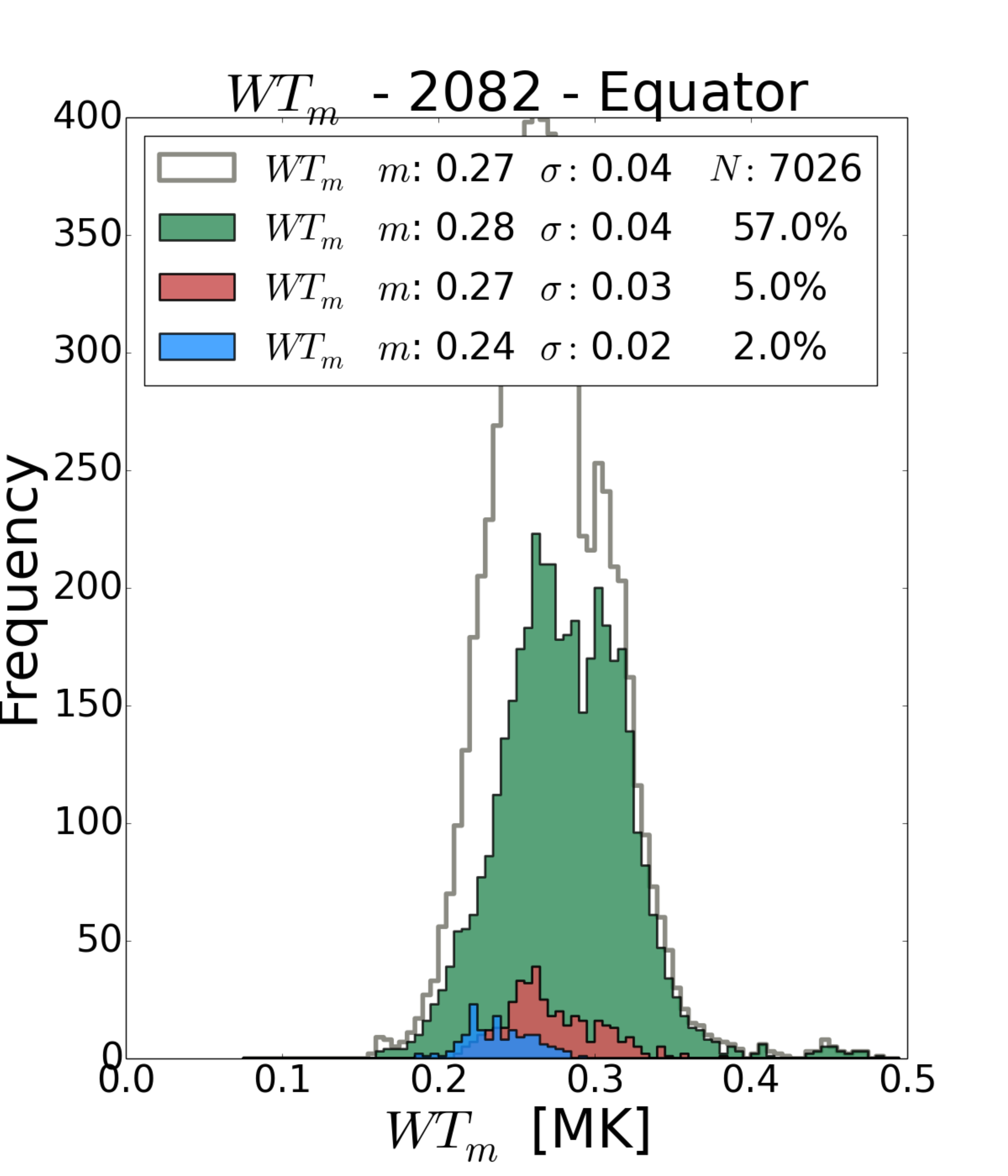}
\includegraphics[width=0.45\textwidth]{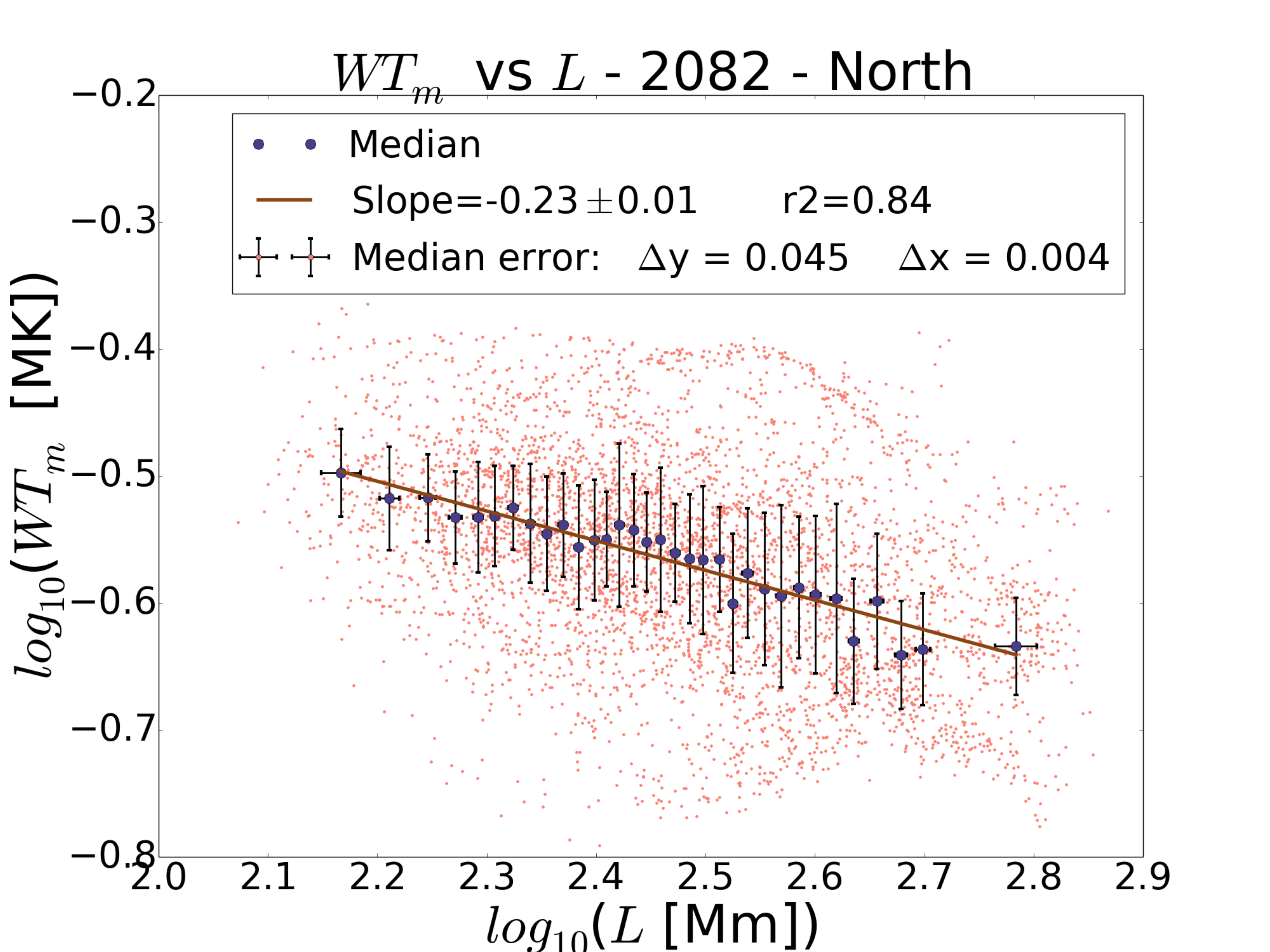}
\includegraphics[width=0.30\textwidth]{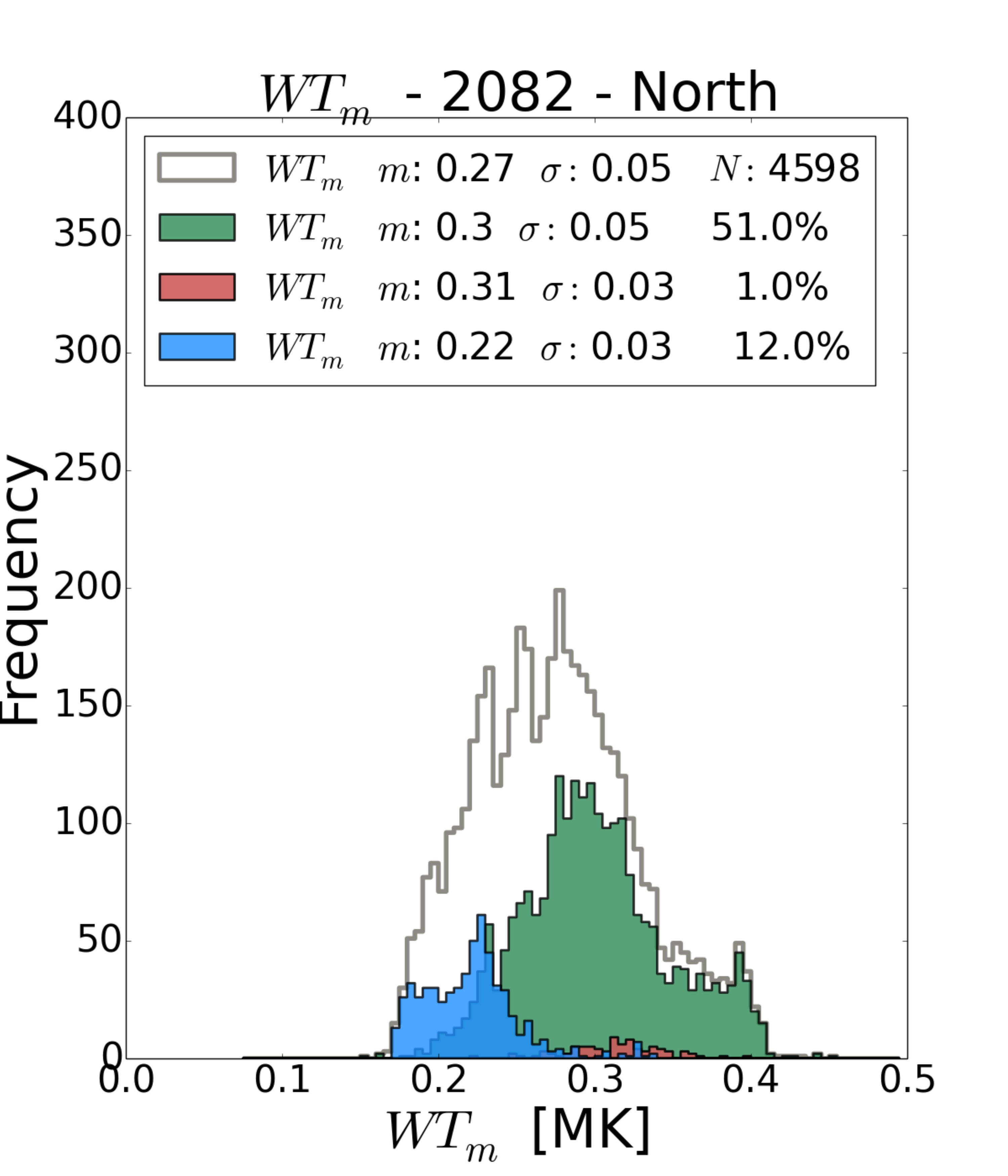}
\includegraphics[width=0.80\textwidth]{Nom.pdf}
\caption{Left panels: Same as Figure \ref{NvsL} for the loop-average temperature distribution width $WT_m$. Right panels: Same as Figure \ref{HistL} for the temperature distribution width $WT_m$.}
\label{WTvsL}
\end{center}
\end{figure*}

The left panels of Fig. \ref{WTvsL} show the loop-average temperature distribution width $WT_m$ as a function of loop length $L$, in each latitude range. We find a similar behavior of the temperature distribution width as a function of length in each region, with slopes of $\approx-0.20$. But when we perform the study of the populations separated by the loop types as described in Section \ref{L}, we find that the loop-average temperature distribution width for up loops is smaller than for down loops in all latitude regions, as it can be seen in the right panels of Fig. \ref{WTvsL}. Given the condition that the temperature variation along the loops must be larger than the loop-average temperature distribution width to be considered non-isothermal, this makes that up loops to tend to present smaller temperature variation along their lengths, so their temperature gradients tend to be smaller than those of the down loops. This will be more evident in Section \ref{Phi} where we analyze the effect of the thermal conduction on the energy balance of the loops. In Fig. \ref{WTvsL} we can also see similar distributions of temperature width for the total sets of loops (gray line), with a median value of $\approx0.27$MK in all latitude regions. 

\subsection{Energy fluxes}\label{Phi}

\begin{figure*}[ht]
\begin{center}
\includegraphics[width=0.28\textwidth]{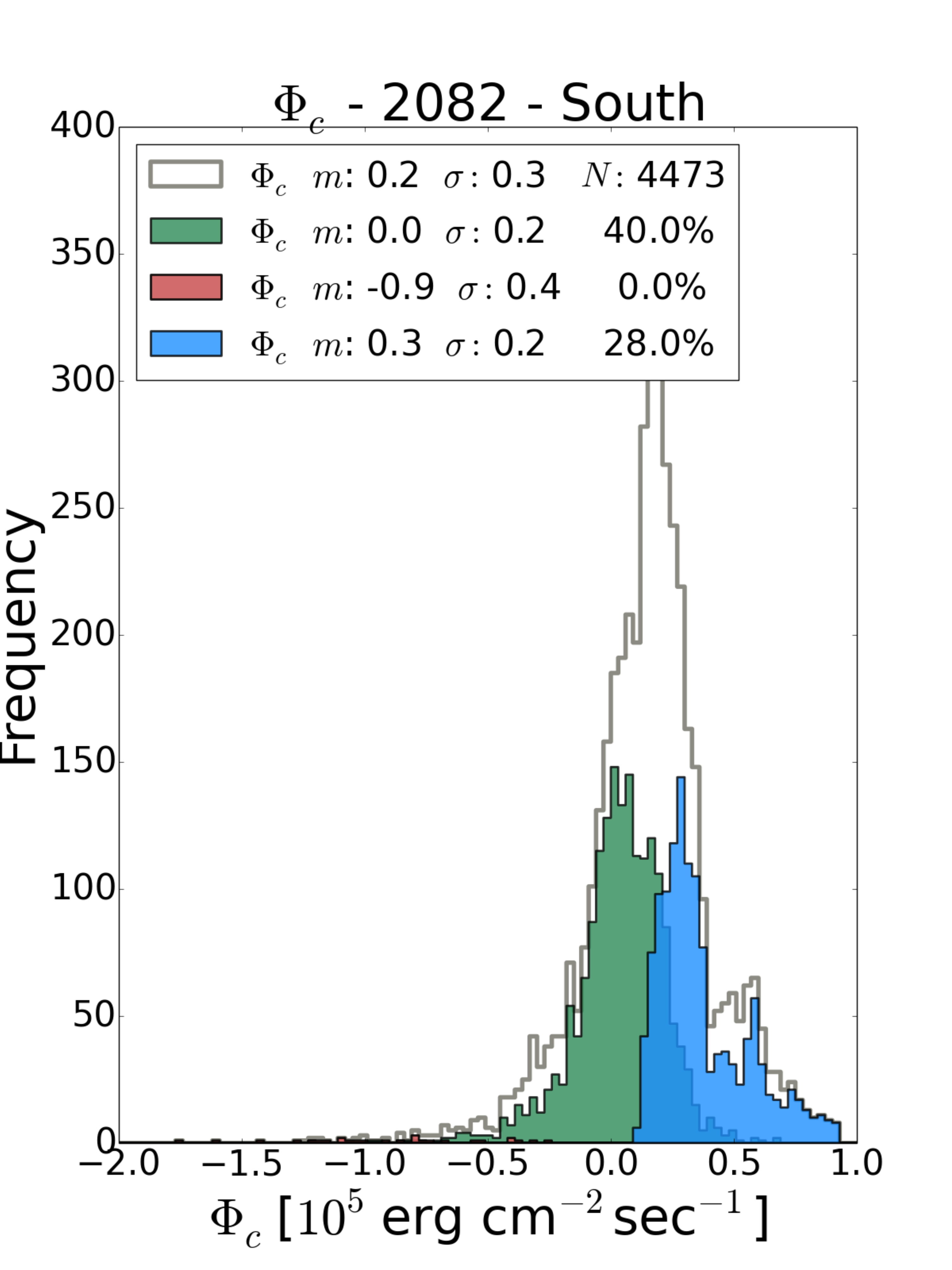}
\includegraphics[width=0.28\textwidth]{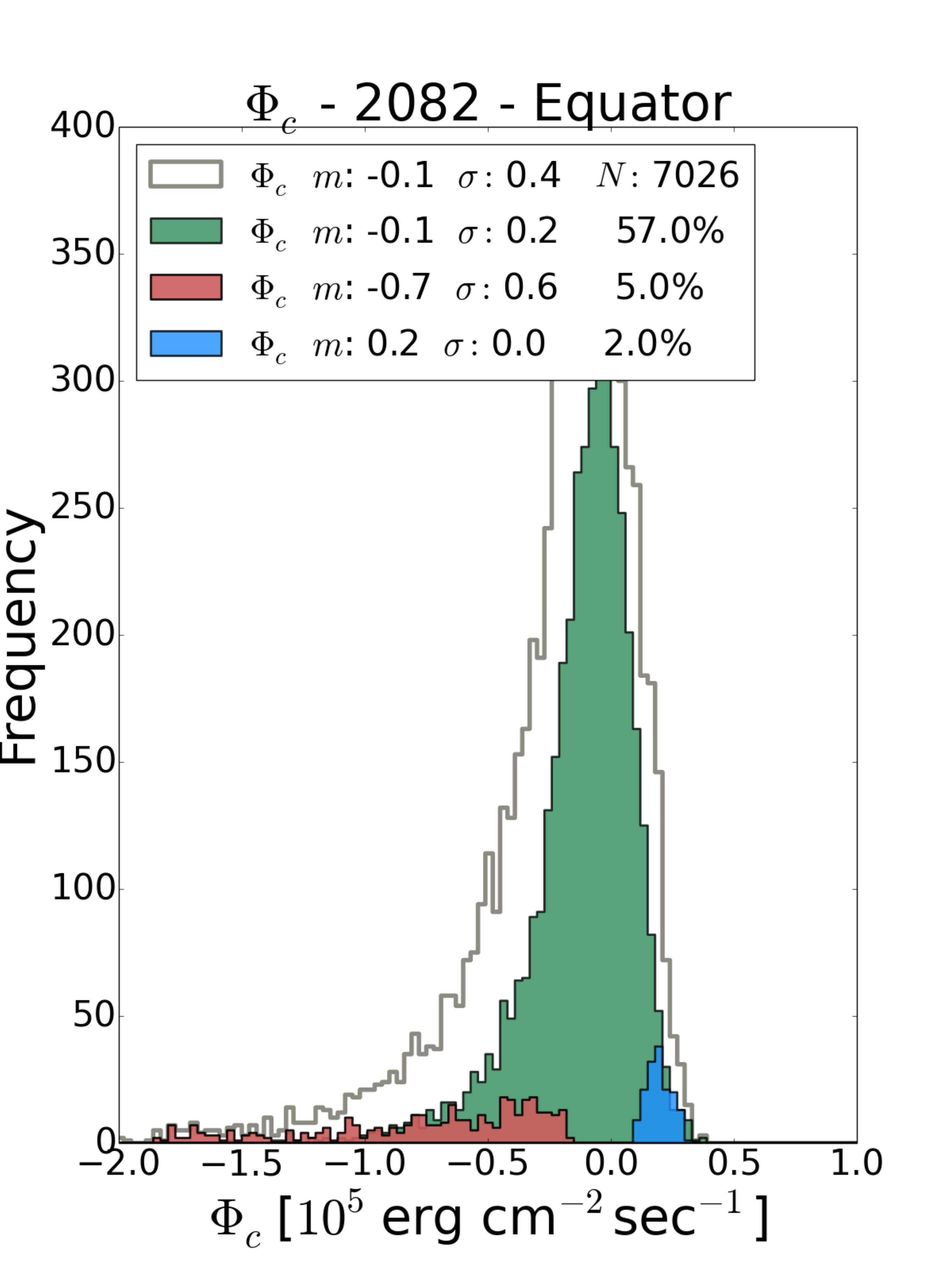}
\includegraphics[width=0.28\textwidth]{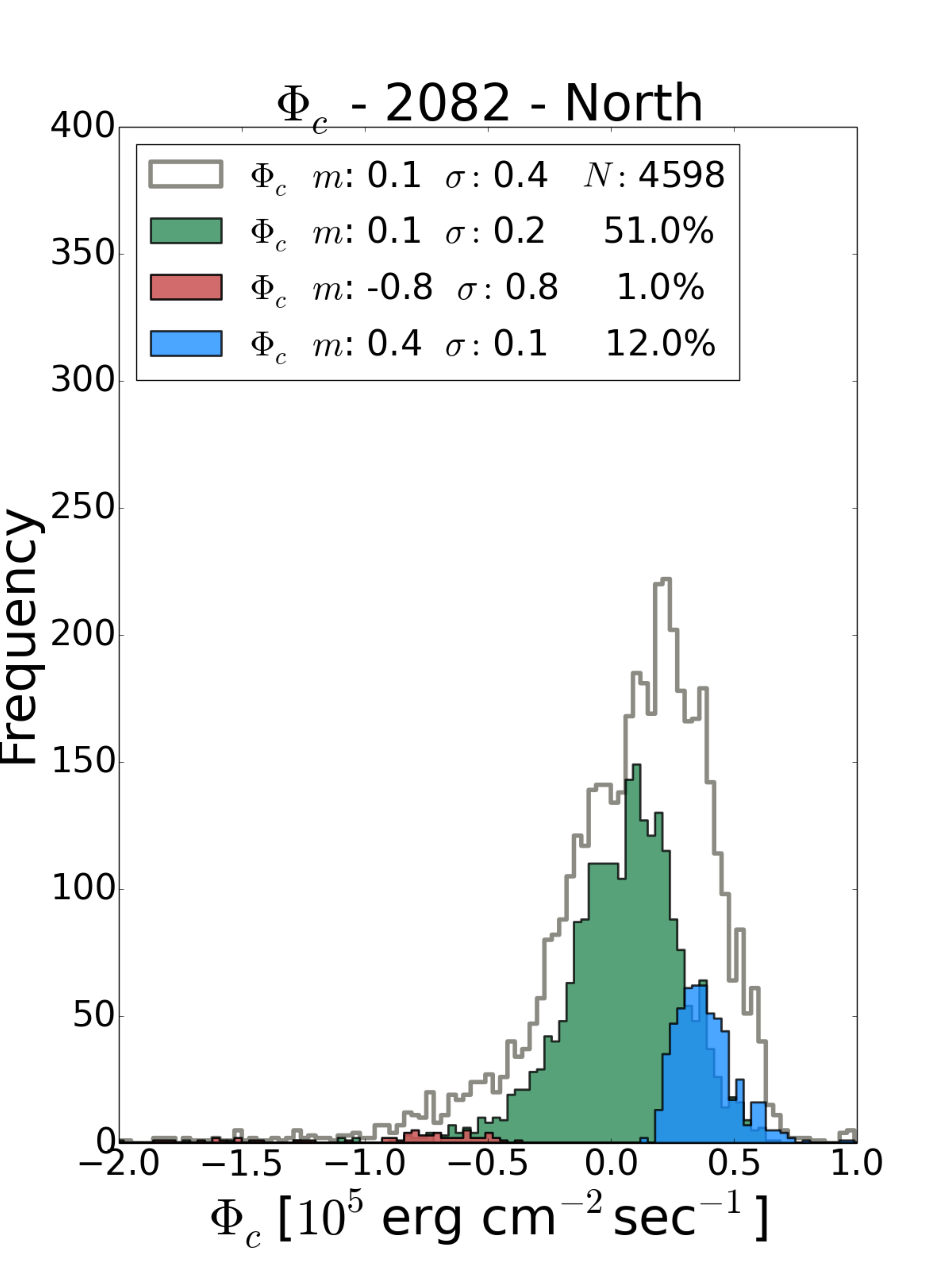}
\includegraphics[width=0.80\textwidth]{Nom.pdf}
\caption{{Same as Figure \ref{HistL} for the loop-integrated conductive flux ($\Phi_c$). We show the percentage of each loop population with respect to the total population.}}
\label{Histc}
\end{center}
\end{figure*}

Fig. \ref{Histc} shows the statistical distribution of the loop-integrated conductive flux, $\Phi_c$, for each latitude region. We observe that for loops in equatorial latitudes, the loop-integrated conductive flux is predominantly negative, while in the other latitudes we see a more symmetrical distribution around zero. This is due to the larger presence of down loops (the red area of the histogram) at the equator \citep[see ][]{maccormack_2017}. Due to the strong dependence of the conductive flux on the temperature gradient along the loops (Eq. \ref{Ec}), we infer that down loops have larger gradients than up loops, producing a distribution of conductive flux displaced towards the negative side, {then an input of heat from the loop footpoints}. We associate this with the temperature distribution width studied in Section \ref{WT}, where we have seen that down loops have larger temperature widths than up loops. Since we used the temperature widths as the threshold to classify up and down loops, the temperature differences between the loop base and the apex tends to be larger for down loops than for up loops. Thus, the gradient is also larger (and negative) producing a negative tail on the conductive flux distribution.

We can also see, in Fig. \ref{Histc}, the percentage of up and down loops with respect to the total number of loops. We see that in the equator $5\%$ of the loops are down loops, while there is no significant population of up loops ($2\%$) and that the isothermal loops are $57\%$ of the total population. In souther latitudes the up loop population is greater than the down loop ($28\%$ and $\approx0\%$ respectively), and the same behavior is observed in the north ($12\%$ of loops are up while there is no significant population of down loops). Notice that with the chosen criteria to classify loops and the wide temperature distributions provided by the tomography, up and down loops are a small fraction of the total population.

{Also, we analyze the behaviour of the conductive flux as a function of length. For north and south latitudes, the conductive flux shows no correlation with length. For equatorial latitudes we find slopes around $\approx-0.9$ with a coefficient of determination of $r^2\approx0.8$. This indicates that negative conductive flux is significantly present in short loops and contributes as an energy input flux at equatorial latitudes.}

\begin{figure*}[ht]
\begin{center}
\includegraphics[width=0.32\textwidth]{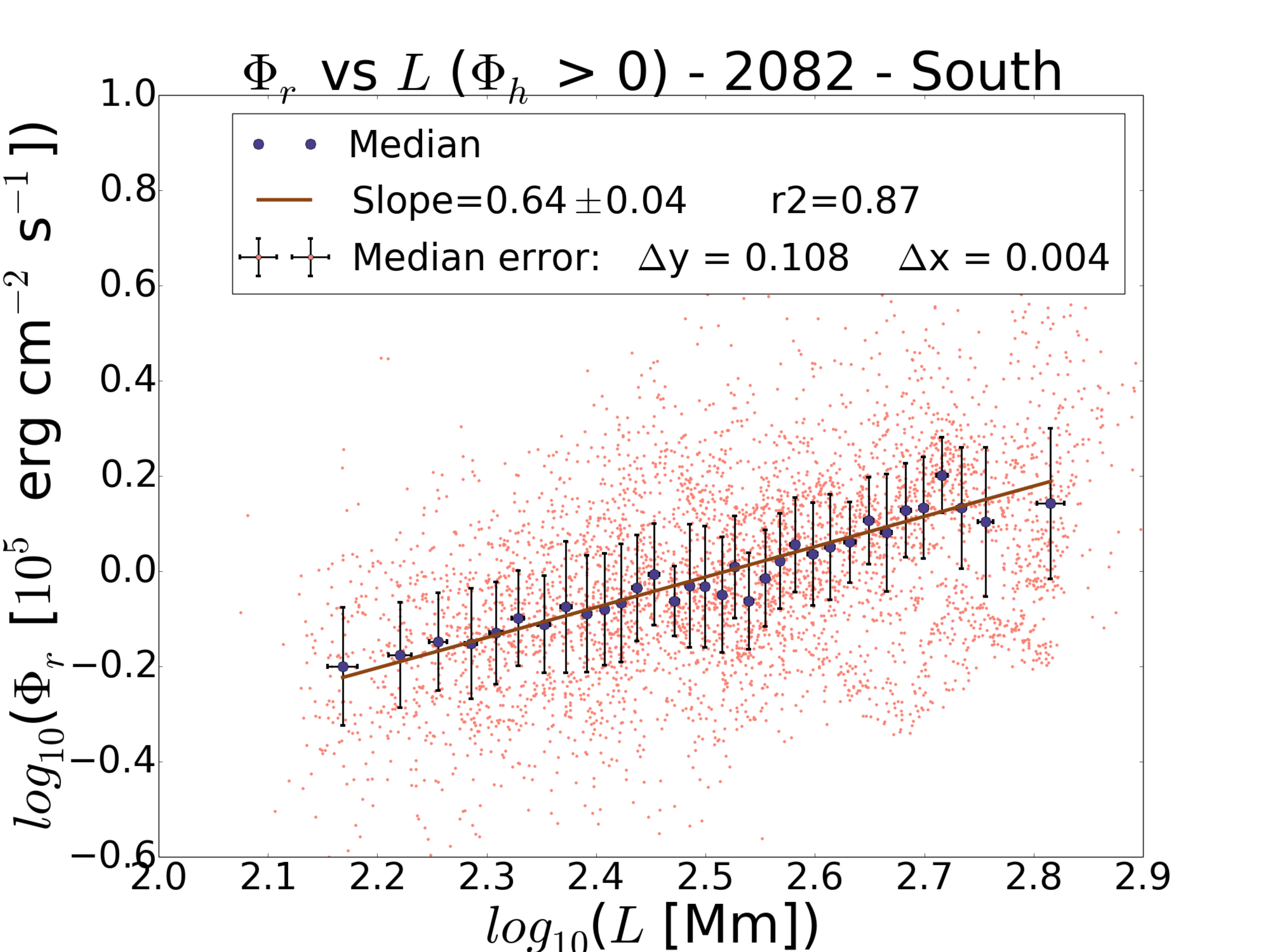}
\includegraphics[width=0.32\textwidth]{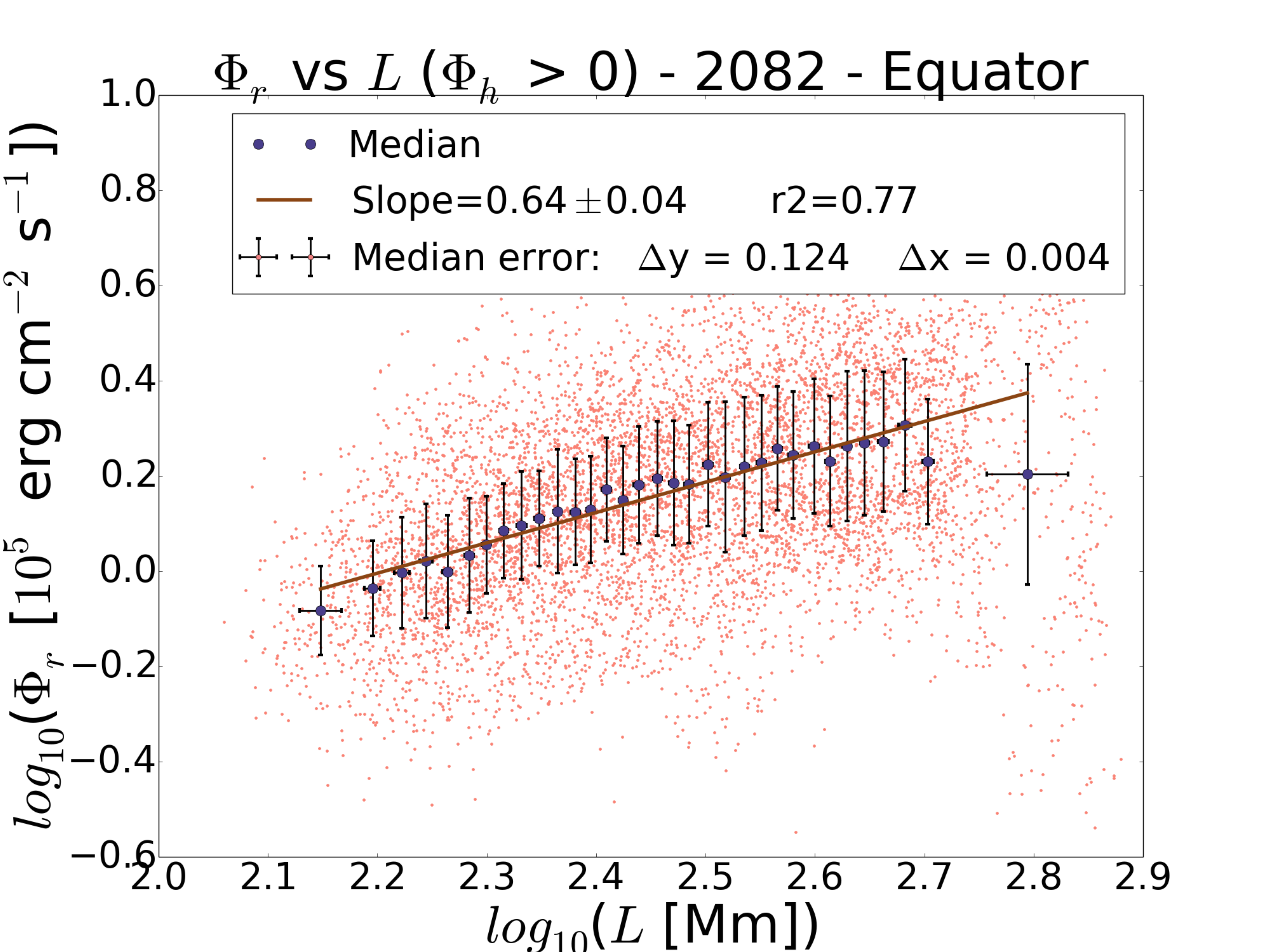}
\includegraphics[width=0.32\textwidth]{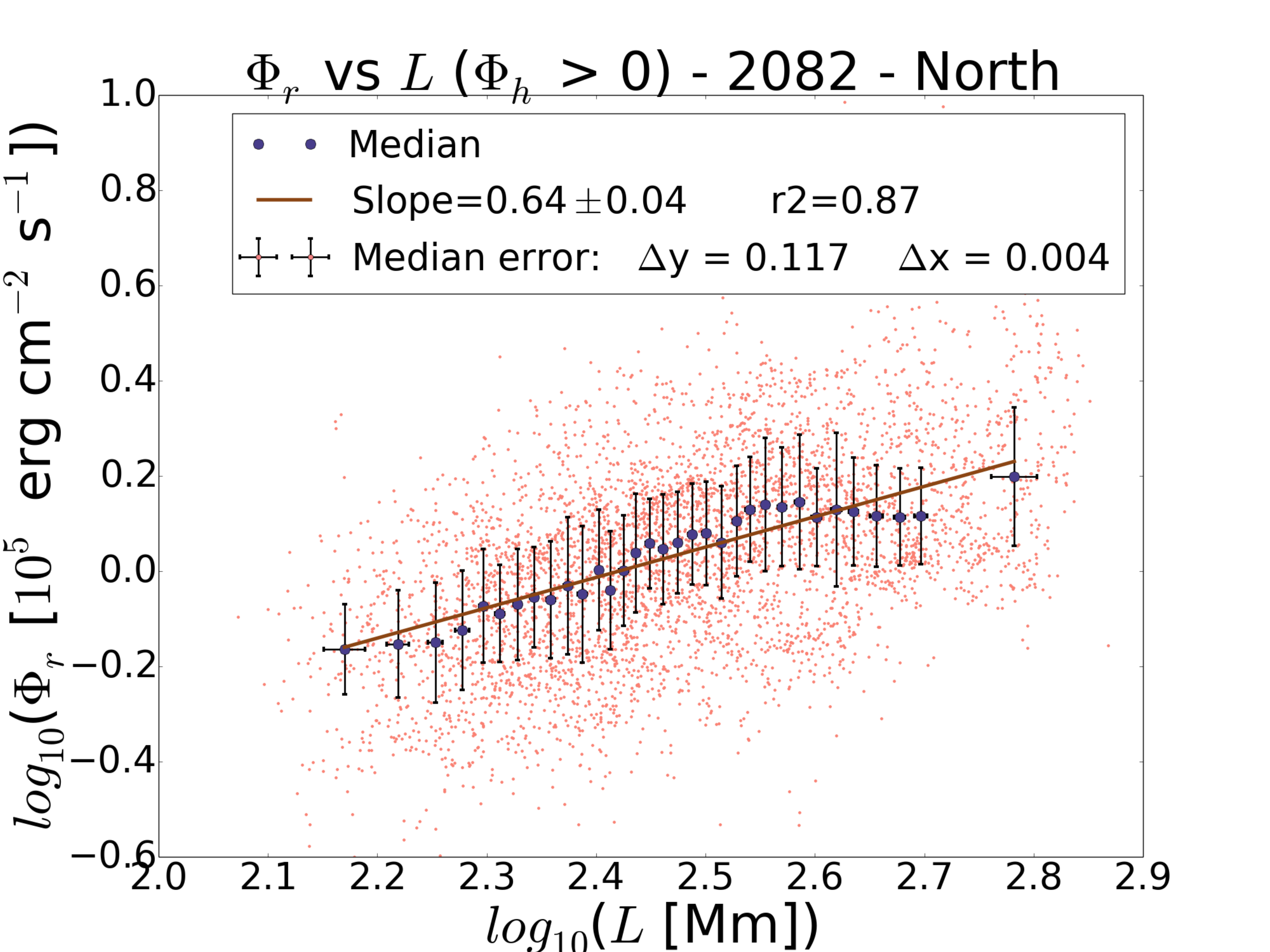}
\caption{Same as Figure \ref{NvsL} for the loop-integrated radiative flux} $\Phi_r$. 
\label{R}
\end{center}
\end{figure*}

In Fig. \ref{R} we plot the loop-integrated radiative flux as a function of length. We obtain similar slopes for the three latitudes ($\approx0.65$). This is expected, since loop-average density has a similar behaviour in the three latitudes and the radiative power is strongly dominated by density (see Eq. \ref{Er} {and \ref{phir2}}). {As a first approximation, we assume a direct relation between square density and length. Since the dependence of the density on loop length is $N_m\approx L^{-0.35}$, as we mentioned before, we would expect to obtain the square of the observed slope. We did not find the expected value because, as we can see in Eq. \ref{phir2}, the magnetic field is also an important variable involved in this relation. In section \ref{Conclusion} we discuss this in more  detail.}

From the balance equation (Eq. \ref{FluxBalance}), all losses should compensate for the gains. But there are some loops for which the loop-integrated conductive flux is so extremely negative (i. e. those down loops with large gradients that we mentioned before) that the loop-integrated radiative flux is not enough to {compensate for such input of energy at the footpoints}. In these loops the loop-integrated energy input flux becomes negative, which of course does not make physical sense. This could be due in part to an underestimation of the radiative flux, considering that we reconstruct the coronal emission using three filter bands, each one with a response at a particular temperature range. It could be possible that part of the plasma is emitting at a temperature outside the detected range, so it is not possible to reconstruct the whole emission profile. {Another possibility is that these loops are in evolution, and what we are seeing is an average of that evolution, since the tomographic technique does not resolve the short timescale dynamics.} We find though that these loops represent only the $\sim2.3\%$ of the total population, so we decided to exclude them from the following analysis. 

\begin{figure*}[ht]
\begin{center}
\includegraphics[width=0.32\textwidth]{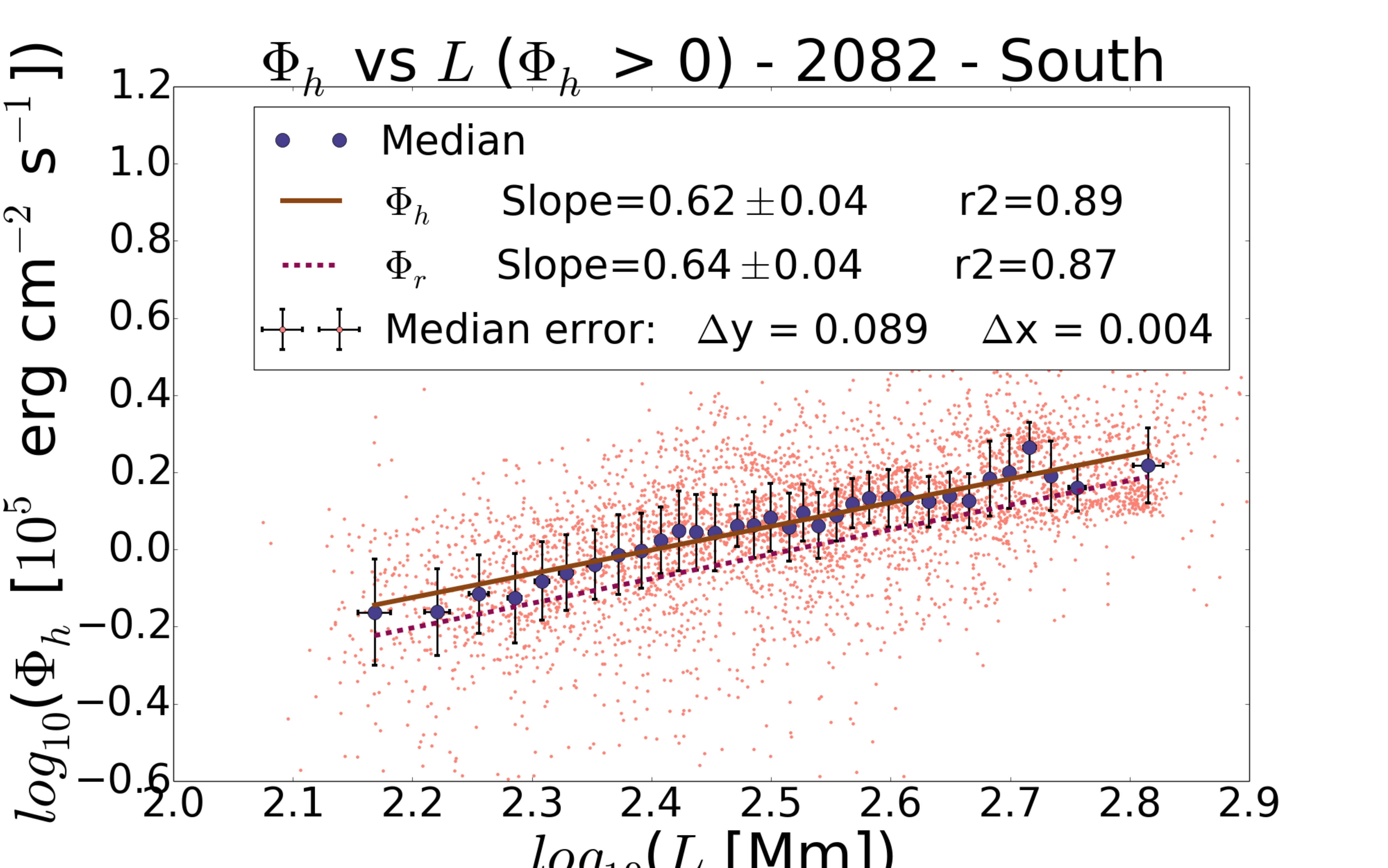}
\includegraphics[width=0.32\textwidth]{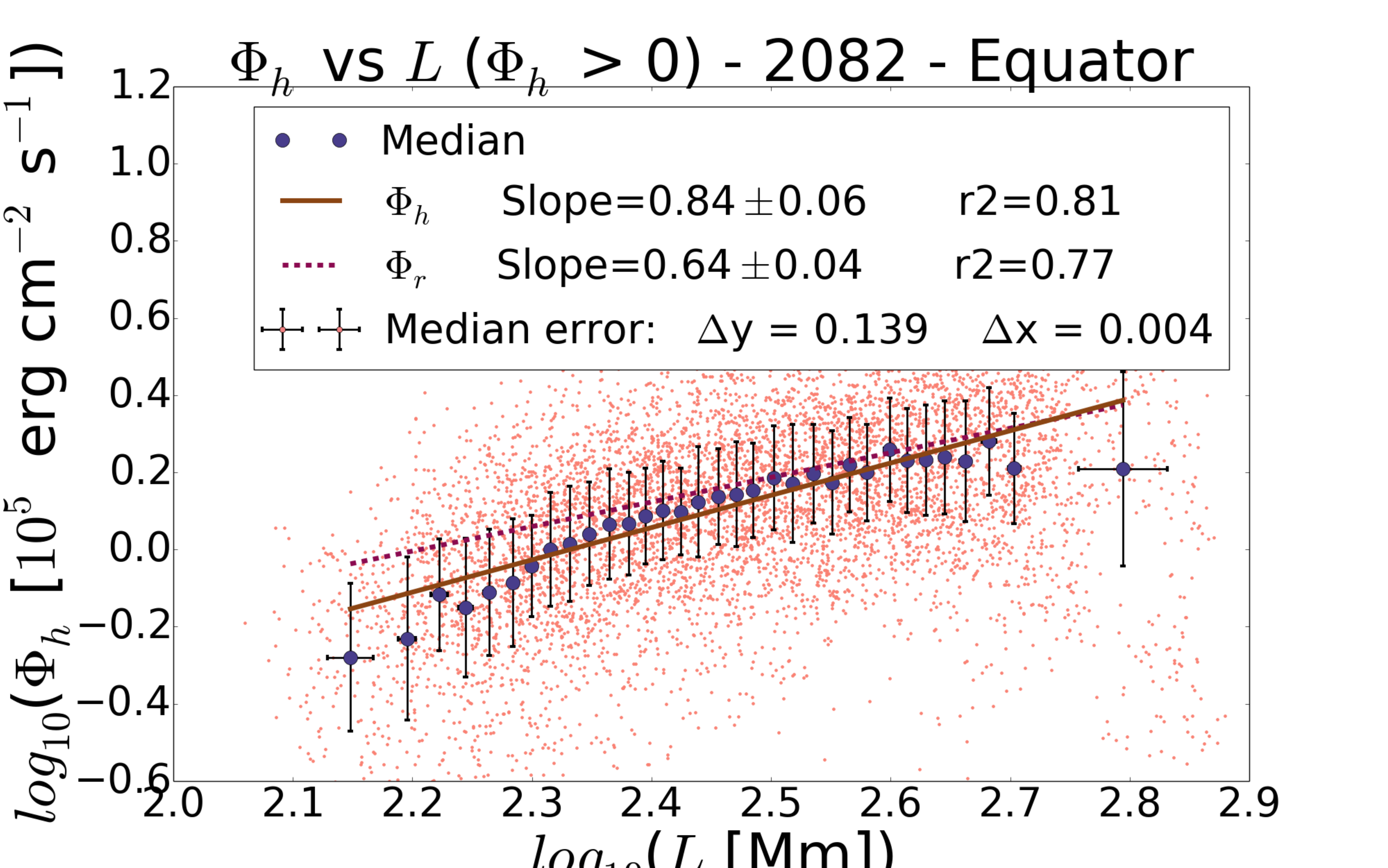}
\includegraphics[width=0.32\textwidth]{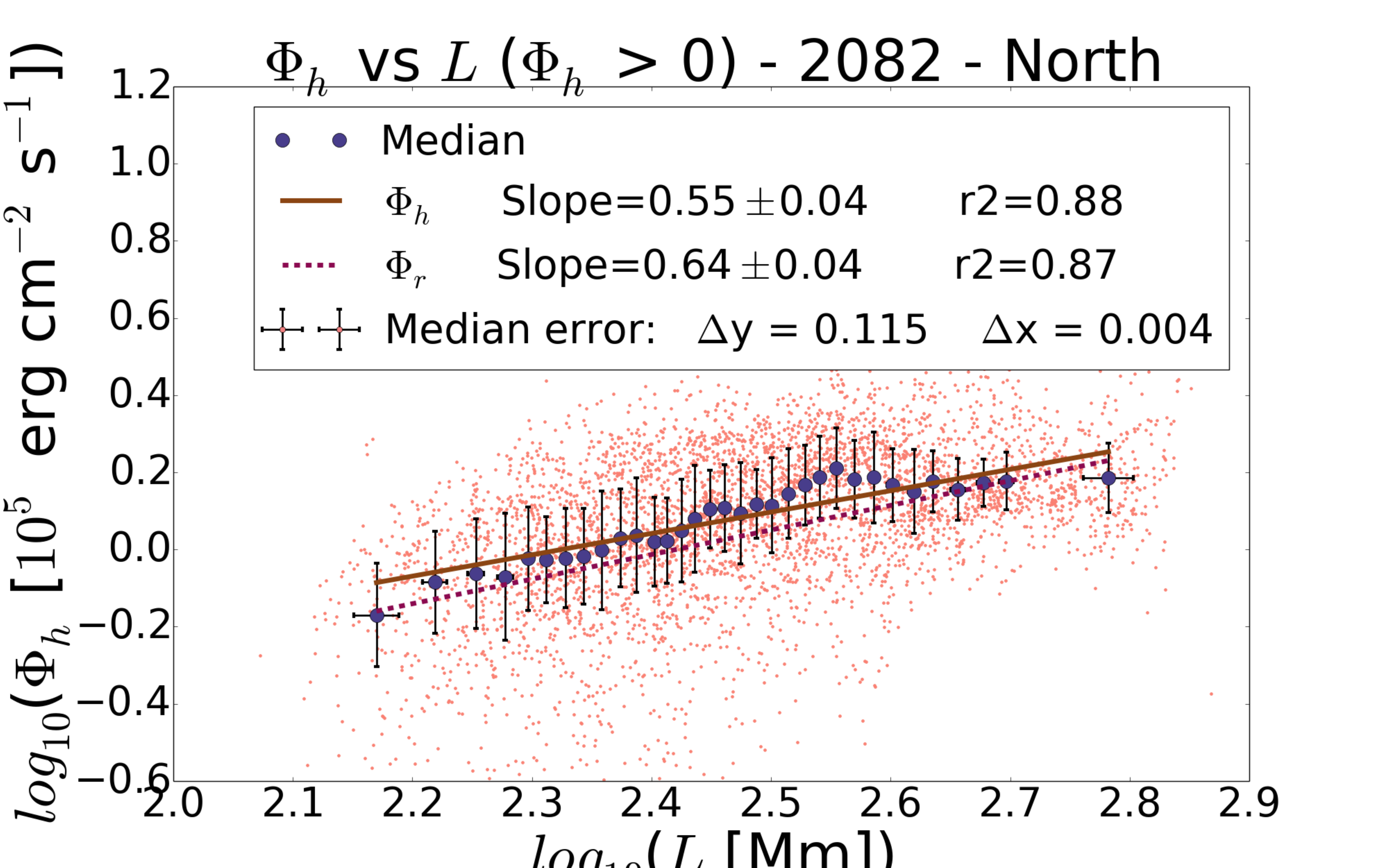}
\caption{Same as Figure \ref{NvsL} for the loop-integrated energy input flux $\Phi_h$. {Dashed-purple line represent the linear fit of the median values of loop-integrated radiative flux in the bins. Both fit results are accompanied by their corresponding coefficient of determination $r^2$.}}
\label{H}
\end{center}
\end{figure*}

Fig. \ref{H} shows the analysis for the loop-integrated energy input flux considering only the positive population. We find similar slopes for the south and north latitudes, and a slope, {within the logarithm plots,} $\approx30\%$ larger in the equatorial latitude. This difference is mainly due of the dominant presence of down loops in the equatorial latitudes. It can be seen that the difference of slope is due to the shorter loops of the equatorial set, which, being dominated by down loops (see Fig. \ref{HistL}, central panel) tend to have a larger negative conductive flux (as shown in the analysis of Figure \ref{Histc}), therefore decreasing the energy flux input and producing a larger slope for the whole set. {Both slopes, loop-integrated radiative flux and energy flux input as a function of length, can be compared in Fig. \ref{H}. It denotes the weak contribution of the conductive flux in southern and northern latitudes and the remarkable presence in the equator.}

\section{Conclusions}\label{Conclusion}

The main aim of this work was to find relations between different plasma parameters along reconstructed quiet-Sun coronal loops. To do that, we used a tomographic technique that provides the 3D distribution of the temperature and electron density in the global corona, obtained from EUV observations in three EUV bands of the STEREO/EUVI telescope integrated along one solar rotation. To reconstruct coronal loops we integrate magnetic field lines from a Potential Source Surface Field (PSSF) model obtained extrapolating the magnetic field from synoptic magnetograms. Combining the DEMT results with the geometrical location of field lines from the magnetic model we obtain loop-averages of temperature and density for each reconstructed loop. We applied the procedure to Carrington rotation 2082.

Our initial approach was to analyze the relation of the thermal properties of the loops with a geometric parameter such as the loop length. Motivated by previous results \citep {nuevo_2013,lloveras_2017}, we separated the loops in three heliographic latitude regions, north (20 to 40 \textdegree), equator (-20 to 20 \textdegree) and south (-40 to -20 \textdegree). Similar length distributions were observed in the three latitude ranges, with a small population of shorter loops ($\approx$150 Mm) at the equator. We also separated loops according to the temperature behaviour along them. Loops for which temperature increases along their length are classified as up, while if the temperature decreases they are classified as down. It was found that down loops are mostly present in equatorial latitudes and, in general, are shorter than up loops, as it has been already observed in previous works \citep {nuevo_2013,lloveras_2017,maccormack_2017}.

For the analysis of the density, we found that, independently of latitude, the loop-average density has a direct relation with length: $N_m \approx L^{-0.35}$. Since radiative flux is directly related to density, as we described in Section \ref{model}, {but also has a strong dependence with the magnetic field \citep{maccormack_2017}, an approximate $\Phi_r \approx N_m^2\,L\,B^{-1}$} relation should be expected. Considering the above relation between $N_m$ and $L$ {and the decreasing relation between the magnetic field and the loop length ($B_m \approx L^{-r}$)}, we would expect then, $\Phi_r\approx L^{0.3+r}$. {In this way, the coefficient related to the magnetic field $r$ should be $\approx 0.3$ which is in the range of coefficients found before. A deeper study of this and other relations with the loop-average magnetic field will be done in the future.} 

No behavior directly linked to the loop length was observed in the loop-average temperature, indicating that within the analyzed coronal heights the temperature is mostly uniform. It is worth reminding though that all expected variations must be in the temperature range within which the instrument has its response. Temperatures outside this range would not be detected. Even so, we observed a dependence of the loop-average temperature on the latitude of the loops. We found that equatorial loops have an average temperature $\approx 15\%$ colder than those in middle latitudes. This could be due to the widespread presence of down loops in the equatorial latitudes, since they tend to be colder than up loops. It is also observed that up loops tend to be less multi-thermal than down loops, which is consistent with the observed relation $WT_m\approx L^{-0.2}$, since down loops are shorter in general than up loops. This has a direct consequence on the temperature gradients used to compute the loop-integrated conductive flux. Down loops have negative gradients that tend to be larger in absolute value than the gradients of up loops, producing larger negative conductive fluxes as we discuss below.

The loop-average magnetic field presents a similar behavior to temperature in relation to latitude. Loops at the equator have a loop-average magnetic field that is $\approx 20\%$ lower than at mid-latitudes. This motivated us to perform a study of the temperature as a function of the magnetic field separating them in latitude bins. In this case, we obtain an approximate relation: $T_m\approx B_m^{0.2}$.

Finally, we analyzed the relation between the loop-integrated energy input flux and the loop length. Since the temperature gradients are very small in the majority of the loops (up and down loops with larger gradients are the minority), the loop-integrated conductive flux is in most cases almost negligible in comparison with the loop-integrated radiative flux. Then, the calculation of the loop-integrated energy input flux is mostly determined by the radiative flux. The case in which the conductive flux has more influence is in the equatorial latitude, where the presence of down loops is more important. In some extreme cases of down loops, the temperature gradient is so negative that the radiative flux is not enough to compensate for the negative conductive flux, producing unphysical negative values of the energy input flux, {since an extra cooling term would be needed for a thermal equilibrium.} {Those loops (representing $\approx 2.3\%$ of the total) were eliminated from the analysis}. Since  down loops are mostly in the short length tail of the distribution, the slope of the loop-integrated energy input flux as a function of loop length is $\approx30\%$ larger in the equatorial latitudes, as we discussed in Section \ref{Phi}.

As we mentioned in Section \ref{intro} a series of scaling laws between loop parameters have been found in studies based on ARs loops. None of the relations found here is fully consistent with those results. This is somehow expected though, because AR loops are shorter ($\approx 50$ Mm, six times smaller than the average of our loops), much denser and hotter, and with magnetic fields in excess of $\approx 100$ G (one or two orders of magnitude larger than the average value found in our loops). While a common approximation in ARs considers loops below the plasma scale-height ($\approx100$ Mm), most of our loops have lengths that far exceed that value. Much of the scaling relations predicted by canonical theoretical estimations \citep[see, e.g.][]{rosner_1978} are based on those physical conditions and approximations. {One important finding by \citet{vesecky_1979}, was that conductive and radiative losses are of the same order of magnitude, while in our study, conductive flux is not significant for the largest part of the loop population. This may be related to the inconsistency of our results with the scaling laws expected in the case of static equilibrium solutions. Therefore, the question arises if our results correspond to loops in equilibrium or they correspond to the averaging of loops which are actually evolving in time.} Some of these considerations have been already discussed by us in \citet{maccormack_2017}, where we compared loop parameters obtained with the tomographic procedure with the results of a 0D hydrodynamic model \citep{klimchuk_2008} that provides mean coronal temperature and density and that included usual assumptions for AR plasmas. We will complement our present results with one-dimensional hydrodynamic models that reproduce the temperature and density profiles of the observed reconstructed loops under physical conditions present in the quiet-Sun corona.


\emph{Acknowledgements} The authors sincerely acknowledge the anonymous reviewers, whose fruitful comments and suggestions enriched and improved the clarity of the article. The authors acknowledge financial support from the Argentinean grants PICT 0221 (ANPCyT), UBACyT 20020130100321, and PIP 2012-01-403 (CONICET).  

\bibliography{MacCormack}
\bibliographystyle{elsart-harv}

\end{document}